\documentclass{pasj01}
\Received{$\langle$reception date$\rangle$}
\Accepted{$\langle$acception date$\rangle$}
\Published{$\langle$publication date$\rangle$}

\usepackage[switch,mathlines]{lineno}

\begin{document}

\title{ACA CO($J$=2--1) Mapping of the Nearest Spiral Galaxy M33. II. Exploring the Evolution of Giant Molecular Clouds}

\author{Ayu \textsc{Konishi}\altaffilmark{1},\footnotemark[*]
Kazuyuki \textsc{Muraoka}\altaffilmark{1},
Kazuki \textsc{Tokuda}\altaffilmark{1,2,3},
Shinji \textsc{Fujita}\altaffilmark{4},
Yasuo \textsc{Fukui}\altaffilmark{5},
Rin I. \textsc{Yamada}\altaffilmark{5},
Fumika \textsc{Demachi}\altaffilmark{5},
Kengo \textsc{Tachihara}\altaffilmark{5},
Masato I. N. \textsc{Kobayashi}\altaffilmark{6},
Nario \textsc{Kuno}\altaffilmark{7,8},
Kisetsu \textsc{Tsuge}\altaffilmark{9,10,11},
Hidetoshi \textsc{Sano}\altaffilmark{10,12},
Rie E. \textsc{Miura}\altaffilmark{13},
Akiko \textsc{Kawamura}\altaffilmark{3},
Toshikazu \textsc{Onishi}\altaffilmark{1}}

\altaffiltext{1}{Department of Physics, Graduate School of Science, Osaka Metropolitan University, 1-1 Gakuen-cho, Naka-ku, Sakai, Osaka 599-8531, Japan}

\altaffiltext{2}{Department of Earth and Planetary Sciences, Faculty of Science, Kyushu University, Nishi-ku, Fukuoka 819-0395, Japan}
\altaffiltext{3}{National Astronomical Observatory of Japan, National Institutes of Natural Sciences, 2-21-1 Osawa, Mitaka, Tokyo 181-8588, Japan}
\altaffiltext{4}{The Institute of Statistical Mathematics, 10-3 Midori-cho, Tachikawa, Tokyo 190-8562, Japan}
\altaffiltext{5}{Department of Physics, Nagoya University, Chikusa-ku, Nagoya 464-8602, Japan}
\altaffiltext{6}{I. Physikalisches Institut, Universit\"{a}t zu K\"{o}ln, Z\"{u}lpicher Str 77, D-50937 K\"{o}ln, Germany}
\altaffiltext{7}{Division of Physics, Faculty of Pure and Applied Sciences, University of Tsukuba, 1-1-1 Tennodai, Tsukuba, Ibaraki 305-8577, Japan}
\altaffiltext{8}{Tomonaga Center for the History of the Universe, University of Tsukuba, Tsukuba, Ibaraki 305-8571, Japan}
\altaffiltext{9}{Institute for Advanced Study, Gifu University, 1-1 Yanagido, Gifu 501-1193, Japan}
\altaffiltext{10}{Faculty of Engineering, Gifu University, 1-1 Yanagido, Gifu 501-1193, Japan}
\altaffiltext{11}{Institute for Advanced Research, Nagoya University, Furo-cho, Chikusa-ku, Nagoya 464-8601, Japan}
\altaffiltext{12}{Center for Space Research and Utilization Promotion (c-SRUP), Gifu University, 1-1 Yanagido, Gifu 501-1193, Japan}
\altaffiltext{13}{Departamento de Fisica Teorica y del Cosmos, Campus de Fuentenueva, Universidad de Granada, E18071-Granada, Spain}

\email{sw23227n@st.omu.ac.jp}

\KeyWords{stars: formation --- galaxies: individual (M33) --- ISM: clouds}

\maketitle
\begin{abstract}
The evolution of giant molecular clouds (GMCs), the main sites of high-mass star formation, is an essential process to unravel the galaxy evolution. Using a GMC catalogue of M33 from ALMA-ACA survey, we classified 848\,GMCs into three types based on the association with H$\,${\sc ii} regions and their H$\alpha$ luminosities \textit{L}\,(H$\alpha$): Type I is associated with no H$\,${\sc ii} regions; Type II with H$\,${\sc ii} regions of \textit{L}\,(H$\alpha$)\,$<$\,10$^{37.5}$\,erg\,s$^{-1}$; and Type III with H$\,${\sc ii} regions of \textit{L}\,(H$\alpha$)\,$\geqq$\,10$^{37.5}$\,erg\,s$^{-1}$. These criteria yield 224 Type I GMCs, 473 Type II GMCs, and 151 Type III GMCs. GMCs show changes in their physical properties according to the types; mass, radius, velocity dispersion, and $^{13}$CO detection rate of GMCs systematically increase from Type I to Type III, and additionally, Type III GMCs are closest to virial equilibrium. Type III GMCs show the highest spatial correlation with clusters younger than 10\,Myr, Type II GMCs moderate correlation, and Type I GMCs are almost uncorrelated. We interpret that these types indicate an evolutionary sequence from Type I to Type II, and then to Type III with timescales of 4\,Myr, 13\,Myr, and 5\,Myr, respectively, indicating the GMC lifetime of 22\,Myr by assuming that Type II GMC has the same timescale as the Large Magellanic Cloud. The evolved GMCs concentrate on the spiral arms, while the younger GMCs are apart from the arm both to the leading and trailing sides. This indicated that GMCs collide with each other by the spiral potential, leading to the compression of GMCs and the triggering of high-mass star formation, which may support the dynamic spiral model. Overall, we suggest that the GMC evolution concept helps illuminate the galaxy evolution, including the spiral arm formation.
\end{abstract}

\section{Introduction}
High-mass stars have a profound effect on the evolution of galaxies by causing supernova explosions which inject enormous kinetic energy and supply heavy elements to interstellar space.
Star-formation-related processes are supposed to change/evolve properties of GMCs as a function of time.
The first observational studies of the time evolution of GMCs began with a CO($J$=1--0) survey toward the Large Magellanic Cloud (LMC) by NANTEN \citep{Fukui08,Fukui99}. 
\citet{Kawamura09} found that the GMCs are classified into three types according to the high-mass star formation activities; Type I shows no signature of high-mass star formation, Type II is associated with only H$\,${\sc ii} region(s), and Type III with both H$\,${\sc ii} region(s) and young stellar clusters (YSCs). 
They argued that the types indicate the evolutionary sequence of GMCs and the GMC lifetime was estimated to be 20 -- 30\,Myr assuming that the timescale in each type is proportional to the number of GMCs. 

Such a classification of GMCs was extended to the spiral galaxy M33, which is a suitable target to study the GMC evolution because its proximity ($D$\,=\,840\,kpc, \cite{Freedman91}) and moderate inclination ($i$\,=\,55$^{\circ}$, \cite{Koch18}) offer a uniform survey of molecular clouds.
\citet{Corbelli17} detected 566 GMCs using CO($J$=2--1) data obtained by IRAM 30\,m at a 50\,pc resolution \citep{Gratier12,Druard14}. They classified the GMCs into different evolutionary stages; inactive, embedded, and exposed star formation by comparing with H$\alpha$, infrared, and far-UV emissions, resulting in the GMC lifetime of 14.2\,Myr. 
Another survey in CO($J$=3--2) covering the smaller fraction of the M33 disk at a 100\,pc resolution is conducted by \citet{Miura12}, yielding the lifetime of GMCs ($>$\,10$^{5}$\,$M_{\odot}$) of 20 -- 40\,Myr.

In the past decade, millimeter-wave interferometers made GMC-scale studies in galaxies within 15 -- 20\,Mpc a realistic task.
\citet{Colombo14} created the large catalogue including 1507 GMCs in the spiral galaxy M51 using CO($J$=1--0) data at a 40\,pc resolution from the PdBI Arcsecond Whirlpool Survey \citep{Schinnerer13}, revealing environmental variations in GMC properties.

More recently, relations between the basic properties of GMCs and star formation have been studied extensively based on the PHANGS-ALMA data \citep{Leroy21}. 
The molecular gas mass detection limit for these surveys is typically 10$^{5}\,M_{\odot}$.
\citet{Schinnerer19} quantiﬁed the relative distribution of CO and H$\alpha$ emission across a large sample of galaxies at a 140\,pc resolution (see also \cite{Pan22}).
\citet{Zakardjian23} reported that feedback by high-mass stars impacts GMC properties with the 30 -- 180\,pc resolution observations.

However, it is still unclear whether the details of GMC evolution, such as the time scale of the GMC and the statistical properties of different evolutionary stages, are universal in the Local Group galaxies.
Of importance in this study is how the evolution of the GMC can be identified in terms of massive star-forming activity (e.g., H$\alpha$ luminosity \textit{L}\,(H$\alpha$) of H$\,${\sc ii} regions).
Furthermore, when discussing the evolution of the GMC, molecular clouds with masses below 10$^{5}$\,$M_{\odot}$ also need to be considered.
Such low-mass clouds, which tend to be quiescent in high-mass star formation, will eventually gain mass to be GMCs through H$\,${\sc i} gas accretion and/or cloud-cloud collisions (see also Section~\ref{D:spiralformation}).
Thus, the low-mass clouds play a crucial role in the process of accumulation of molecular clouds.

Thus, for the key step to understand the GMC evolution, 
it is necessary to establish a method to quantify the GMC evolution and to examine detailed physical properties of GMCs at different star formation phases and their environmental dependencies using a observation data with high spatial resolution and mass dynamic range across an entire galaxy.
Recently, \citet{Muraoka23} conducted a CO($J$=2--1) multi-line survey with a 30\,pc resolution in M33 by the 7\,m antenna of the Atacama Compact Array (ACA). 
The ACA survey revealed the molecular gas distribution in detail across the entire disk, and promoted an extensive study of clumpy, small, and low-mass (10$^{3}$ -- 10$^{4}$\,$M_{\odot}$) molecular clouds. 

In this paper, we attempt to apply a type classification of GMCs using the ACA CO($J$=2--1) data with the highest resolution and sensitivity in M33, where we use only \textit{L}\,(H$\alpha$) of H$\,${\sc ii} regions as an indicator of the GMC type because a spatially-resolved-cluster-catalogue of M33 based on optical band which is comparable to that of the LMC \citep{Bica96} is unfortunately not available.
We demonstrate that H$\alpha$ emission works as a good tracer of the GMC evolution as well as that of the high-mass star formation.
In addition, we reveal the nature of each type of GMC; not only basic properties (mass, size, velocity dispersion) but also previously undisclosed detailed physical properties (size-linewidth relation, virial parameter, dense gas fraction) and spatial distributions.

This paper is organized as follows; the data used are described in Section~\ref{sec:data}, and Sections~\ref{R:type} and \ref{R:gmcprop} introduce the classification method and the GMC properties, respectively.
Section~\ref{R:ysc} presents the comparison between GMCs and YSCs to confirm the validity of the classification based on \textit{L}\,(H$\alpha$).
Note that individual clusters could not be resolved in M33 unlike that of the LMC due to the insufficient spatial resolution of infrared data used for the identification of YSCs.
In Section~\ref{D:lifetime}, we estimate the GMC lifetime and compare it with the previous studies in M33.
In Section~\ref{sec:dis}, we discuss the evolutionary process of GMCs, and the effect of galactic dynamics on the evolution and lifetime of GMCs.
Finally, we summarize the main results in Section~\ref{sec:sum}.

\section{Data} \label{sec:data}
\subsection{The ACA CO($J$=2--1) data and GMC catalogue}\label{da:CO}
Observations toward M33 were carried out in $^{12}$CO($J$=2--1), $^{13}$CO($J$=2--1), and C$^{18}$O($J$=2--1) by the ALMA-ACA 7\,m antennas at a spatial resolution of 30\,pc (project code: 2017.1.00901.S; 2018.A.00058.S; 2019.1.01182.S).
The ACA $^{12}$CO($J$=2--1) data cube was combined with IRAM 30\,m data \citep{Druard14} to compensate for extended molecular-line emission (hereafter referred to as $``$ACA+IRAM $^{12}$CO($J$=2--1)$"$ data). 
The detailed ACA data analyses were described in \citet{Muraoka23}.
We used the molecular cloud catalogue obtained by the ACA+IRAM $^{12}$CO($J$=2--1) data \citep{Muraoka23}.
This catalogue contains 848 clouds (hereafter $``$GMCs$"$ in this paper) with a mass range from 6.7\,$\times\,10^{3}\,M_{\odot}$ and 2.6\,$\times\,10^{6}\,M_{\odot}$.

\subsection{The H$\alpha$ data}\label{da:ha}
We have used continuum-subtracted H$\alpha$ data \citep{Hoopes00,Hoopes01} and star cluster catalogues \citep{Meulenaer15,Corbelli17} in order to investigate the high-mass star formation activity within GMCs.
The H$\alpha$ data was obtained by the Kitt Peak National Observatory (KPNO) 2.1\,m telescope.
Its angular resolution is about 3\farcs0.
\citet{Hoopes00} reported that the observation field is 1.15\,deg $\times\,$ 1.15\,deg and the total \textit{L}\,(H$\alpha$) is $\sim$3 $\times\,$10$^{40}$\,erg\,s$^{-1}$ in this area. 
Based on this information, we converted the unit of FITS data to erg\,s$^{-1}$.

\subsection{The star cluster catalogues}\label{da:sc}
\citet{Meulenaer15} assembled a star cluster catalogue by combining three catalogues (\cite{San10}; Ma \,\yearcite{Ma12}, \yearcite{Ma13}; \cite{Fan14})
and derived the physical parameters toward 910 star clusters in optical ($\it{UBVRI}$) with near-infrared ($\it{JHK}$) photometric systems.
In addition, \citet{Corbelli17} catalogued 630 YSC candidates, which are identified using Spitzer 24\,$\mu$m image, and whose masses and ages are derived via multi-wavelength spectral energy distribution (SED) fitting by \citet{Sharma11}.
The \citet{Corbelli17} catalogue is more sensitive to embedded young star clusters than the optically-based-catalogue by \citet{Meulenaer15}.
The cluster-age ranges of \citet{Meulenaer15} and \citet{Corbelli17} catalogues are 10$^{6.6}$ -- 10$^{10.1}$\,yr and 10$^{6.0}$ -- 10$^{7.1}$\,yr, respectively. 
The details of these catalogues are given in the Appendix~\ref{app:clct}.
Note that these catalogues could not spatially resolve individual clusters as opposed to the LMC clusters catalogue \citep{Bica96} used in \citet{Kawamura09} because of the insufficient spatial resolution of data used for the identification and cataloguing of the clusters.
Thus each catalogued cluster in M33 is actually composed of multiple clusters.

\section{Type classification of GMCs} \label{R:type}
\subsection{GMC classification criteria}\label{R:typeclass}
CO survey on GMC scales toward external galaxies discovered that there are large variations in star formation activity between GMCs; e.g., the association with star clusters and/or H$\,${\sc ii} regions (e.g., \cite{Fukui99,Kawamura09,Onodera10,Schruba10,Miura12,Corbelli17}). 
In the LMC, \citet{Yamaguchi01} found that \textit{L}\,(H$\alpha$) of the H$\,${\sc ii} regions which are associated with GMCs increase as GMCs evolve.
\textit{L}\,(H$\alpha$) within Type II GMCs which are associated with only H$\,${\sc ii} regions are less luminous than 10$^{37.5}$erg\,s$^{-1}$, while those within Type III GMCs with both H$\,${\sc ii} regions and YSCs are almost more luminous than 10$^{37.5}$erg\,s$^{-1}$. 
This suggests that most of the H$\,${\sc ii} regions associated with YSCs are more luminous than 10$^{37.5}$erg\,s$^{-1}$. The \textit{L}\,(H$\alpha$) of 10$^{37.5}$erg\,s$^{-1}$ is a typical brightness of an H$\,${\sc ii} region ionized by either a single O-type star or several B-type stars.
Therefore, we consider that Type III GMCs, which have the potential to host young clusters, can be distinguished based only on their \textit{L}\,(H$\alpha$).

On the basis of this speculation, we try classifying GMCs into following the three types based on the association with H$\,${\sc ii} regions and their \textit{L}\,(H$\alpha$);
\begin{description}
\setlength{\itemsep}{1pt}
   \item[Type I$\,\colon$] GMCs associated with no H$\,${\sc ii} regions \\
   \, \qquad (inactive high-mass star-formation)
   \item[Type II$\,\colon$] GMCs associated with H$\,${\sc ii} regions of \\ 
   \, \qquad \textit{L}\,(H$\alpha$)\,$<$ 10$^{37.5}$erg\,s$^{-1}$ 
   \item[Type III$\,\colon$] GMCs associated with H$\,${\sc ii} regions of \\ 
   \, \qquad \textit{L}\,(H$\alpha$)\,$\geqq$ 10$^{37.5}$erg\,s$^{-1}$ 
\end{description}
We use the same term, I, II, and III, as \citet{Kawamura09}. Although our definitions of the classifications are different from those in \citet{Kawamura09}, we consider our types to be equivalent to the previous ones.

\subsection{Identification of H$\,${\sc ii} regions}\label{R:id}
As a first step, we performed identification of H$\,${\sc ii} regions in the H$\alpha$ map by using \texttt{astrodendro} algorithm \citep{Rosolowsky08}.
This tool allows us to extract and analyze hierarchical structures in astronomical data including H$\,${\sc ii} regions (e.g., \cite{Zaragoza15, Weilbacher18, Della20, McLeod21, Cosens22}).
Identified structures by \texttt{astrodendro} are composed of leaves (the local maxima that have no substructures), branches (intermediate structures that contain leaves and other branches), and trunks (structures at the lowest level that contain all other structures). 
Here, we define leaves as H$\,${\sc ii} regions.

There are three input parameters in \texttt{astrodendro}, \texttt{min\_value}: the minimum intensity value to be considered, \texttt{min\_delta}: the minimum height for the leaf structure to be considered as an independent entity, \texttt{min\_npix}: the minimum number of two-dimensional pixels in the R.A. and Decl.\,axes. 
The \texttt{astrodendro} package is designed to allow users to determine structures subjectively.
Since our primary goal in this study is to identify molecular clouds associated with H$\;${\sc ii} regions and compare them to those studied in the LMC \citep{Kawamura09}, we chose parameters that would roughly reproduce the typical size and flux of H$\;${\sc ii} regions in the LMC \citep{Kennicutt86}.
The \texttt{min\_value} and \texttt{min\_delta} are set to about 2\,$\times\,$10$^{34}$\,erg\,s$^{-1}$ ($\sim$6\,$\sigma$) and 1\,$\times\,$10$^{34}$\,erg\,s$^{-1}$ ($\sim$3\,$\sigma$), respectively.
The \texttt{min\_npix} of 25\,pix (1\,pix = 1\farcs5) corresponds to about three times as large as the telescope beam size.
Eventually, the \texttt{astrodendro} analysis identified 794 structures as leaves.
The minimum \textit{L}\,(H$\alpha$) is $\lesssim$\,10$^{36}$\,erg\,s$^{-1}$, which is consistent with that used for the LMC study \citep{Kennicutt86,Kawamura09}. 
The uncertainty in \textit{L}\,(H$\alpha$) derived using a bootstrap method \citep{Rosolowsky06} is 20\,$\%$ at most, 
and about 80\,$\%$ of H$\,${\sc ii} regions exhibit uncertainties less than 5\,$\%$.
Regarding the size distribution, most entities range between several tens pc to 100\,pc with uncertainties of around 5\,$\%$, which is also consistent with the LMC catalogue (see also the identified structure map and frequency distribution of \textit{L}\,(H$\alpha$) and radius for H$\;${\sc ii} regions provided in the Appendix~\ref{app:idHa}).
Strictly speaking, accurately determining the boundary of an individual H$\;${\sc ii} region is difficult due to its intrinsically diffuse nature. However, the total flux of each H$\;${\sc ii} region is determined by the most intense part, i.e., leaves in our analysis, of the relevant hierarchical structure.

\subsection{Determination of GMC -- H$\,${\sc ii} regions association and their H$\alpha$ luminosities}\label{R:typeclass}
We determined the association of GMCs and H$\,${\sc ii} regions if the spatial extent of a GMC overlaps with the H$\alpha$ emitting region identified by \texttt{astrodendro}. Note that multiple GMCs can be associated with the same H$\,${\sc ii} region such as the NGC~604 region (see also Section~\ref{R:indivregions}).

\begin{figure}[hbtp]
 \begin{center}
  \includegraphics[width=80mm]{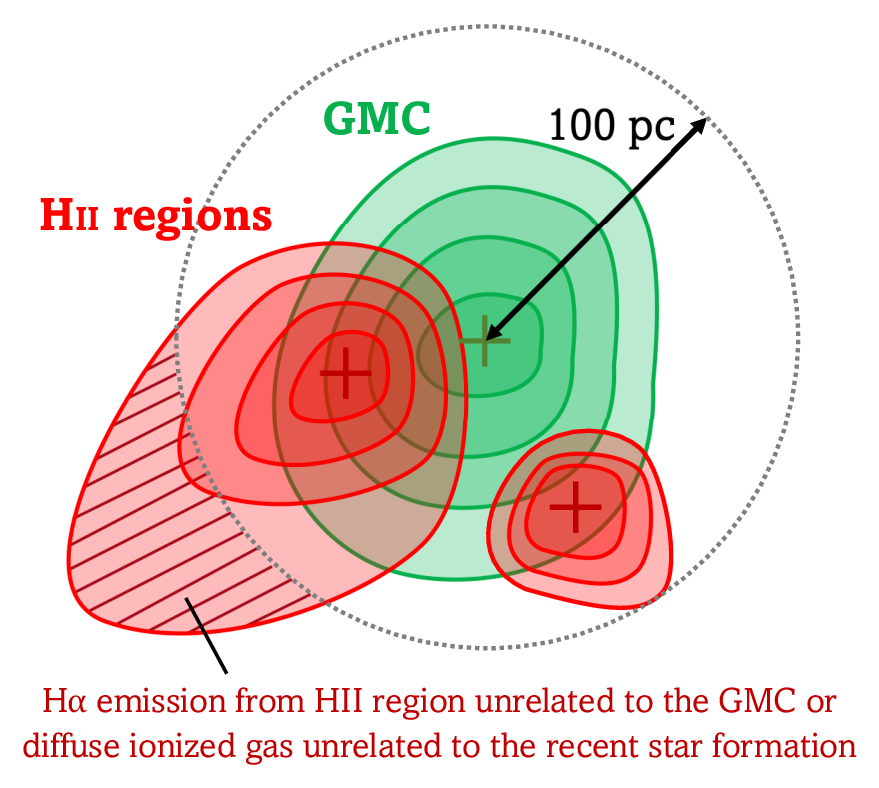}
 \end{center}
 \caption{Schematic view of matched GMC -- H$\,${\sc ii} regions. The association of a GMC and H$\,${\sc ii} regions are determined if the spatial extent of a GMC overlaps with the H$\alpha$ emitting region identified by \texttt{astrodendro}. We defined the \textit{L}\,(H$\alpha$) for the classification criteria as the sum of the H$\alpha$ luminosities of H$\,${\sc ii} regions within a 100\,pc radius from the $^{12}$CO intensity peak of the GMC in order to minimize the contribution of H$\alpha$ emission unrelated to the GMC.
}\label{fig:matchedGMCHa}
\end{figure}

The  \textit{L}\,(H$\alpha$) for the criteria of the type classification are defined as the sum of the H$\alpha$ luminosities of H$\,${\sc ii} regions within a radius of 100\,pc from the $^{12}$CO intensity peak of the GMC (Figure~\ref{fig:matchedGMCHa}).
H$\,${\sc ii} regions are considered to be formed within nearby GMCs as their parent cloud.
In the LMC, high-mass star formation traced by H$\,${\sc ii} regions and YSC with an age younger than 10\,Myr are spatially correlated within $\sim$100\,pc of the molecular clouds \citep{Kawamura09}.
Thus we set the radius as 100\,pc.
This is to minimize the contribution of H$\alpha$ emission from H$\,${\sc ii} regions unrelated to the GMC when the GMC is associated with a giant H$\,${\sc ii} region, such as NGC~604 or NGC~595, and also to minimize the effect of surrounding diffuse ionized gas unrelated to the recent star formation in the GMC.
Here, we note the impact of the change in metallicity on the stellar feedback and separation between CO and H$\alpha$ emission.
CO is more easily photodissociated by UV radiation of high-mass stars in low-metallicity environments, which yields that a large portion of molecular gas become CO-dark gas (e.g., \cite{Maloney88,Bolatto99,Tokuda21}).
The metallicity of M33 is roughly half-solar (12 + log (O/H) = 8.36 $\pm$ 0.04; \cite{Simon08}) with a shallow metallicity gradient. This is similar to that of the LMC (12 + log (O/H) = 8.37 $\pm$ 0.09; \cite{Kurt98}).
Thus, we expect that the influence of stellar feedback is also similar between M33 and the LMC.

\begin{figure*}[htbp]
 \begin{center}
  \includegraphics[width=160mm]{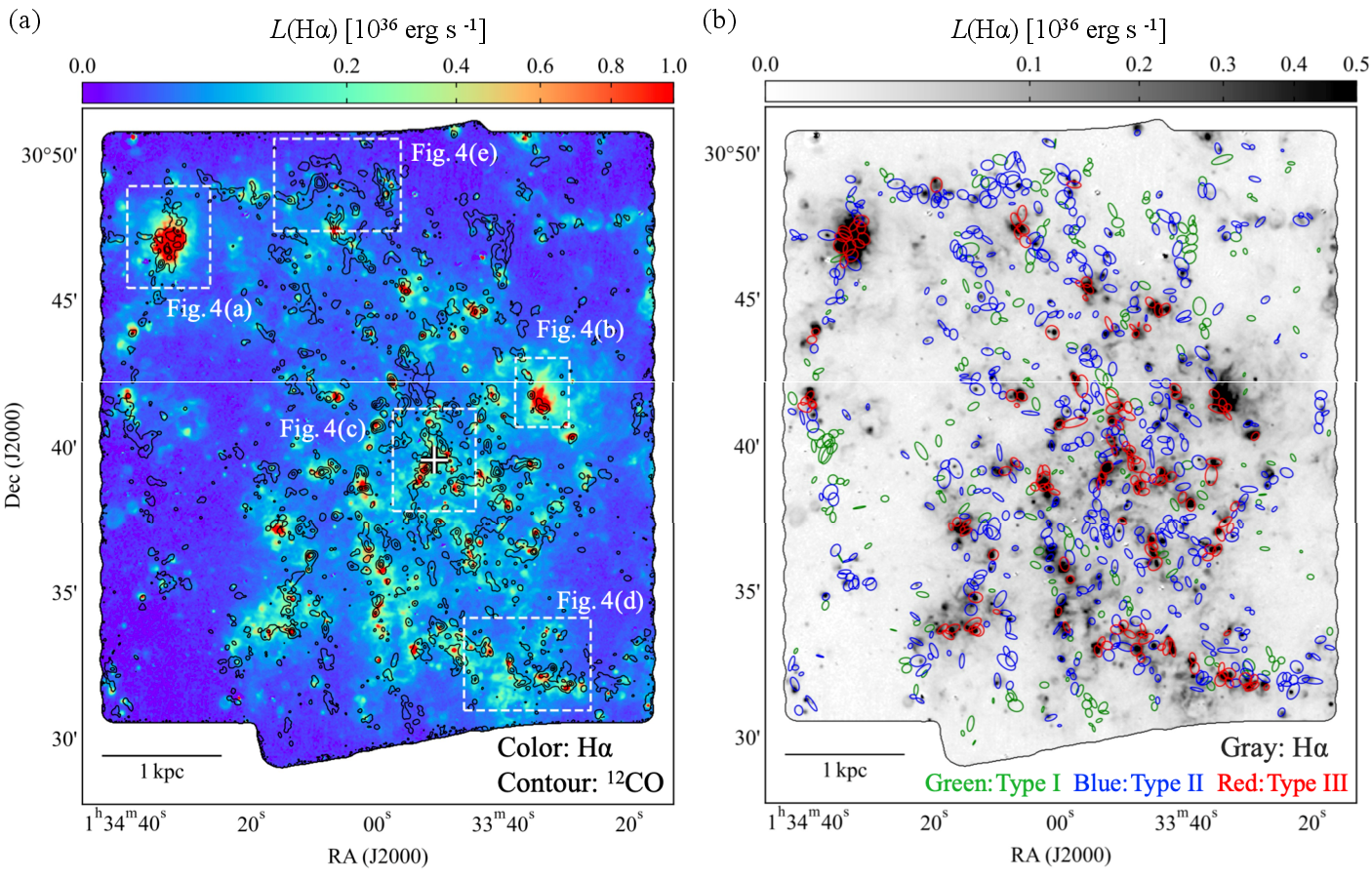}
 \end{center}
 \caption{(a) $^{12}$CO($J$=2--1) emission obtained by the ACA 7\,m array + IRAM 30\,m \citep{Muraoka23} on the H$\alpha$ image \citep{Hoopes00,Hoopes01}, which is cut to fit the ACA observation field.
 Black contours represent $^{12}$CO($J$=2--1) peak temperature and the contour levels are 0.25, 1, and 2\,K, respectively. 
Boxes with dashed lines indicate the areas in Figure~\ref{fig:zoominGMCs}~(a) -- (e). The white cross represents the galaxy's infrared center of (R.A.(J2000), Decl.(J2000)) = (1$^{\rm h}$33$^{\rm m}$50$^{\rm s}$.9, 30$^{\circ}$39\arcmin37\arcsec) \citep{Skrutskie06}.
(b) Spatial distribution of each type of GMC on the H$\alpha$ image. Green, blue, and red ellipses show the extrapolated and beam-deconvolved size and orientations of Type I, II, and III GMCs, respectively. The gray scale shows the same H$\alpha$ image as (a).
}\label{fig:typeGMCsonHa}
\end{figure*}

Then, if the multiple H$\,${\sc ii} regions are associated with a single GMC, we estimated their total H$\alpha$ luminosities.
These criteria yield 224 (26\,$\%$) Type I GMCs, 473 (56\,$\%$) Type II GMCs, and 151 (18\,$\%$) Type III GMCs.
Figure~\ref{fig:typeGMCsonHa}~(a) shows the spatial distributions of H$\alpha$ and CO emissions, while the distribution of each type of GMC is presented in Figure~\ref{fig:typeGMCsonHa}~(b).

\subsection{CO -- H$\alpha$ luminosities correlations between GMC and H$\,${\sc ii} region}\label{R:typeclass}
\begin{figure*}[htbp]
 \begin{center}
  \includegraphics[width=160mm]{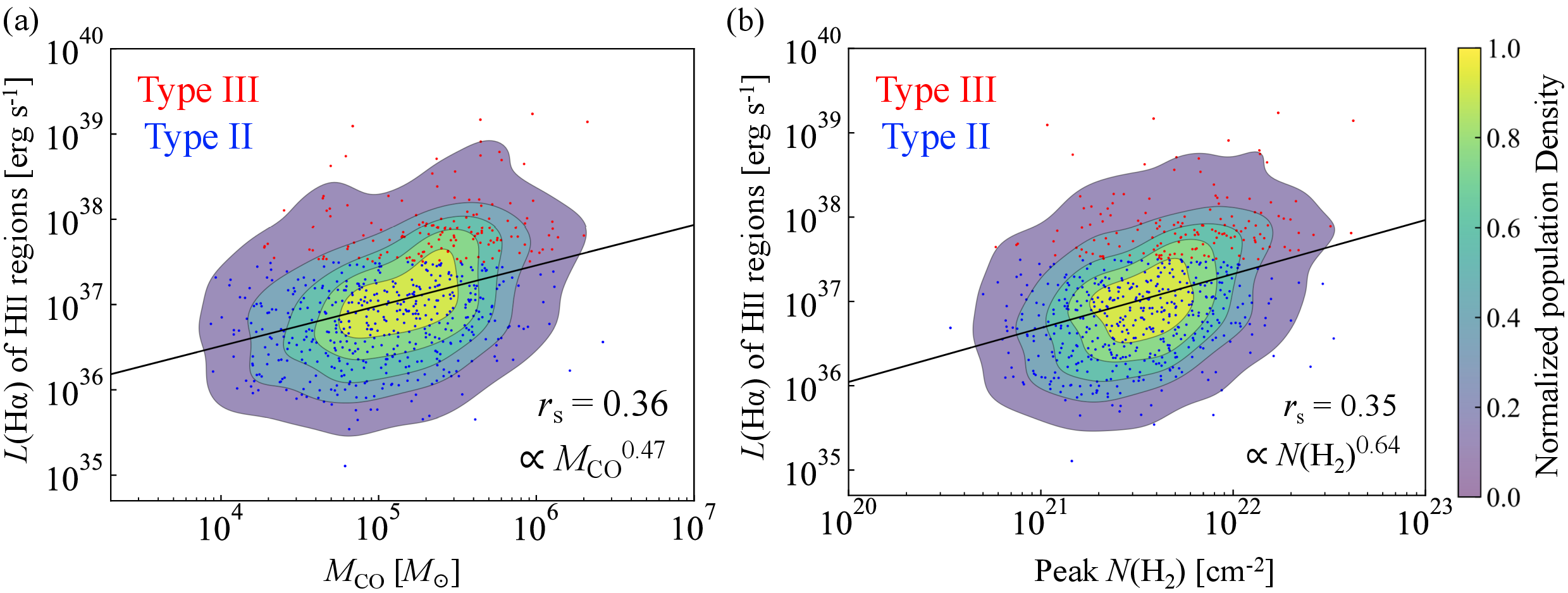}
 \end{center}
 \caption{Scatter plot of correlations (a) between $M_{\rm CO}$ and \textit{L}\,(H$\alpha$), and (b) between peak \textit{N}($\textrm{H}_{2}$) and \textit{L}\,(H$\alpha$).
Blue and red points represent the properties of Type II and Type III GMCs, respectively.
Black lines show the least-square fits for all data points. 
The Spearman's rank correlation coefficients of plots (a) and (b) are 0.36 and 0.35, respectively.
}\label{fig:coHaintensity}
\end{figure*}

A lot of GMCs ($>$\,70\,$\%$) are associated with H$\,${\sc ii} regions, which indicates that GMCs are well spatially correlated with H$\,${\sc ii} regions across the molecular-gas disk of M33.
Figure~\ref{fig:coHaintensity} represents the \textit{L}\,(H$\alpha$) for H$\,${\sc ii} regions associated with the GMC as functions of the CO luminosity mass $M_{\rm CO}$ and the peak column density \textit{N}($\textrm{H}_{2}$) for GMCs. 
$M_{\rm CO}$ is derived from $^{12}$CO($J$=2--1) luminosity assuming $^{12}$CO($J$=2--1)/$^{12}$CO($J$=1--0) ratio $R_{21}$ of 0.6 \citep{Muraoka23} and a CO-to-H$_{2}$ conversion factor $X_{\rm CO}$ of 4.0\,$\times$ 10$^{20}$\,cm$^{-2}$(K km\,s$^{-1}$)$^{-1}$\citep{Gratier17},
which is calculated as 
\begin{eqnarray}
\frac{M_{\rm CO}}{M_{\odot}} = \frac{4.35 \,\, X_{\rm CO}}{2.0 \times 10^{20}\,{\rm cm}^{-2} ({\rm K\,km\,s}^{-1})^{-1}} \, \frac{L_{\rm CO}}{{\rm K\,km\,s}^{-1}\,{\rm pc}^2} \, R_{21}^{-1}. \nonumber \\%
\end{eqnarray}

The relations of $M_{\rm CO}$ and peak \textit{N}($\textrm{H}_{2}$) with \textit{L}\,(H$\alpha$) show a large scatter, but weakly correlate; \textit{L}\,(H$\alpha$) $\propto$ $M_{\rm CO}$$^{0.47}$ and \textit{L}\,(H$\alpha$) $\propto$ \textit{N}($\textrm{H}_{2}$)$^{0.64}$.
The Spearman's rank correlation coefficient between \textit{L}\,(H$\alpha$) and $M_{\rm CO}$ is 0.36, and that between \textit{L}\,(H$\alpha$) and peak \textit{N}($\textrm{H}_{2}$) is 0.35,
although the dynamic range of the vertical axis ($\sim$\,4 orders of magnitude) is different from that of the horizontal axis ($<$\,3 orders of magnitude).
These results indicate that the star formation activity of GMCs is weakly correlated with the amount of molecular gas.

\begin{figure*}[htbp]
 \begin{center}
  \includegraphics[width=153mm]{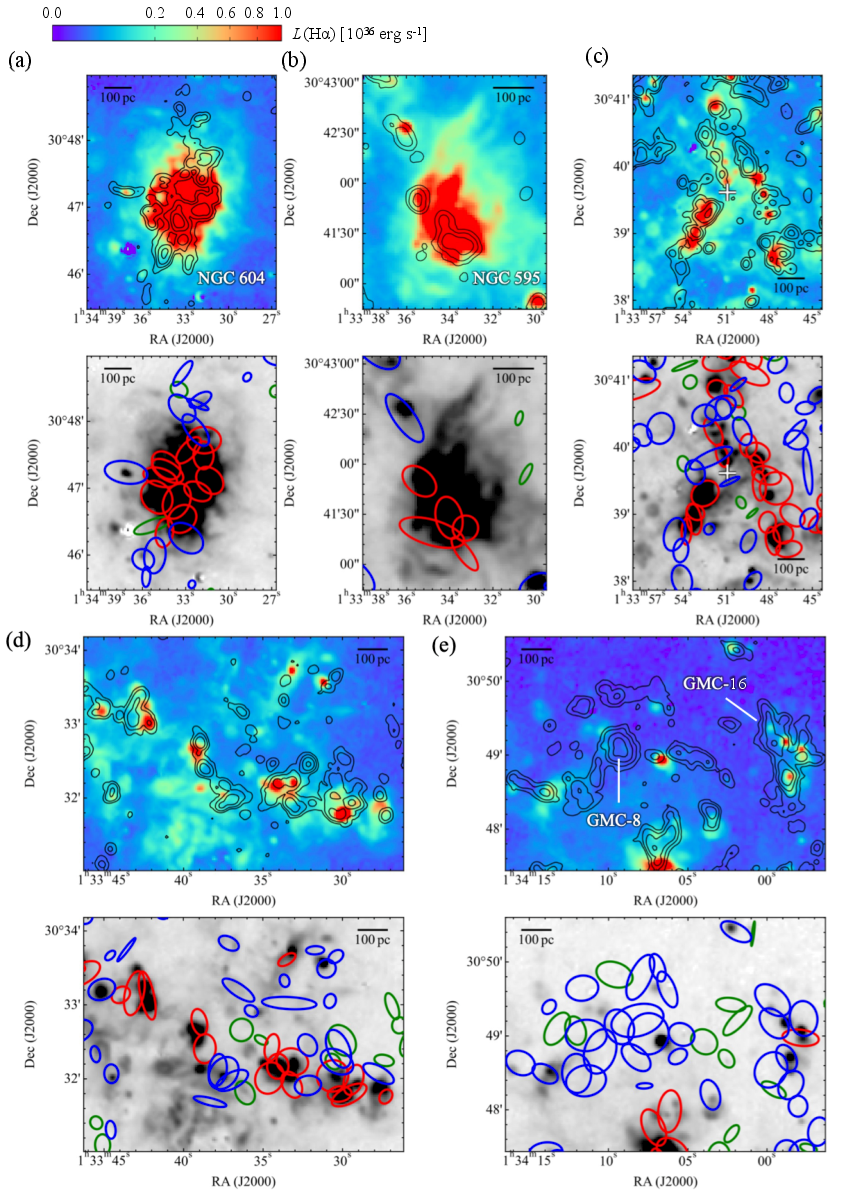}
 \end{center}
 \caption{Upper panels: $^{12}$CO($J$=2--1) peak temperature (contours) on the continuum-subtracted H$\alpha$ luminosity for (a) NGC~604 which is the most luminous H$\;${\sc ii} region in M33, (b) the second most luminous H$\;${\sc ii} region NGC~595, (c) the central region in M33, (d) the south-eastern part of M33 where H$\;${\sc ii} regions align continuously, and (e) some of the most massive ($\sim$10$^{6}$\,$M_{\odot}$) GMCs in M33; GMC-8, GMC-16.
The contour levels are 0.25, 0.5, 1, and 2\,K, respectively. 
Lower panels: The gray scale shows the same H$\alpha$ image as upper panels.
Green, blue, and red ellipses show Type I, II, and III GMCs, respectively. The white cross in (c) represents the infrared center of M33.
}\label{fig:zoominGMCs}
\end{figure*}

\subsection{Individual star-forming regions}\label{R:indivregions}
In Figure~\ref{fig:zoominGMCs}, we present zoomed-in views of some individual star-forming regions.
Figure~\ref{fig:zoominGMCs}~(a) shows NGC~604, which is the most luminous H$\;${\sc ii} region in M33.
NGC~604 has at least 200 O-type stars with ages ranging from 3 to 5\,Myr \citep{Hunter96}.
We identified the H$\,${\sc ii} region with the \textit{L}\,(H$\alpha$) of $\sim$10$^{39.3}$\,erg s$^{-1}$ in the NGC~604 region by \texttt{astrodendro}.
12 Type III GMCs are associated with NGC~604 and their total GMC mass is $\sim\,8.0\,\times\,10^{6}\,M_{\odot}$.
NGC~604 is one of the most massive GMC-association in the Local Group.

Figure~\ref{fig:zoominGMCs}~(b) shows the second most luminous H$\;${\sc ii} region NGC~595, which contains $\sim$250 OB-type stars with an age of 4.5\,Myr \citep{Malumuth96} and is considered to be more evolved than NGC~604 from the comparison of the OB-type stars' age, a Wolf-Rayet to O-type star ratio, and molecular photo-dissociation between the NGC~595 and NGC~604 regions \citep{Wilson95}.
The \textit{L}\,(H$\alpha$) measured by \texttt{astrodendro} is $\sim$10$^{39.0}$\,erg\,s$^{-1}$ and four Type III GMCs with a total GMC mass of $\sim\,1.3\,\times\,10^{6}\,M_{\odot}$ are associated with NGC~595. 
As opposed to NGC~604, the amount of molecular gas in NGC~595 shows no outstanding difference from that in other star-forming regions in spite of its star formation activity.
It is probably because molecular gas is now dissipating due to stellar feedback \citep{Wilson95}.

In the central region of M33 (Figure~\ref{fig:zoominGMCs}~c), many Type III GMCs are distributed surrounding the galaxy's infrared center indicated by a cross.
Figure~\ref{fig:zoominGMCs}~(d) shows the south-eastern part of M33.
There are luminous H$\,${\sc ii} regions around the arm, where Type II and Type III GMCs are predominantly distributed.

Figure~\ref{fig:zoominGMCs}~(e) shows GMC-8 and GMC-16 which are some of the most massive ($>$ 10$^{6}$\,$M_{\odot}$) GMCs in M33. ALMA observed them at 1\,pc resolution, revealing the detailed molecular gas distributions and physical properties in detail \citep{Tokuda20,Kondo21}.
GMC-8 shows inactive high-mass star formation in spite of its rich molecular reservoir \citep{Miura12,Kondo21} and is classified into Type II in this work. GMC-16 have filamentary structures elongated in the north–south direction (see also \cite{Tokuda20}) and is associated with some H$\,${\sc ii} regions.

\section{The nature of the GMCs at different star formation phases}\label{R:gmcprop}
To reveal the physical condition of GMCs leading to high-mass star formation, it is necessary to investigate the basic properties, scaling relations, dense gas fraction, and spatial distributions of GMCs at various star formation phases within a galaxy.
In this section, we examine the nature of the three types of GMCs. 

The physical properties of GMCs are summarized in Table~\ref{table:clsum}.
Although the details of the derivation of the GMC properties are found in \citet{Muraoka23},
we provide their brief summary here. 
\citet{Muraoka23} identified GMCs by using \texttt{PYCPROPS} algorithm \citep{Rosolowsky21}.
\texttt{PYCPROPS} provides extrapolated parameters of identified structures such as the standard deviation of the major ($\sigma_{\rm maj}$), minor ($\sigma_{\rm min}$), and velocity ($\sigma_{v, {\rm ext}}$) axes, and integrated intensity in order to measure physical properties without sensitivity bias.
This extrapolation method is described in \citet{Rosolowsky06} and \citet{Rosolowsky21}. 
Then we obtained a deconvolved GMC radius and velocity dispersion by the beam size $\sigma_{\rm beam}$ and the velocity channel width $\Delta V_{\rm chan}$, respectively.
The effective radius $R$ is calculated as follows:
\begin{eqnarray}
R = 1.91 \sqrt{(\sigma_{\rm maj}^2 - \sigma_{\rm beam}^2)^{0.5} (\sigma_{\rm min}^2 - \sigma_{\rm beam}^2)^{0.5}}.
\end{eqnarray}
The deconvolved standard deviation of the velocity $\sigma_{v}$ is calculated as follows:
\begin{eqnarray}
\sigma_v = \sqrt{\sigma_{v, {\rm ext}}^2 - \frac{\sigma_{v, {\rm chan}}^2}{2 \pi}},
\end{eqnarray}
where $\sigma_{v, {\rm chan}} = \Delta V_{\rm chan}/(2 \sqrt{2\,{\rm ln}\,2})$ ($\Delta V_{\rm chan}$ = 0.7\,km\,s$^{-1}$).

\subsection{Basic properties}\label{R:bprop}
\begin{figure*}[htbp]
 \begin{center}
  \includegraphics[width=160mm]{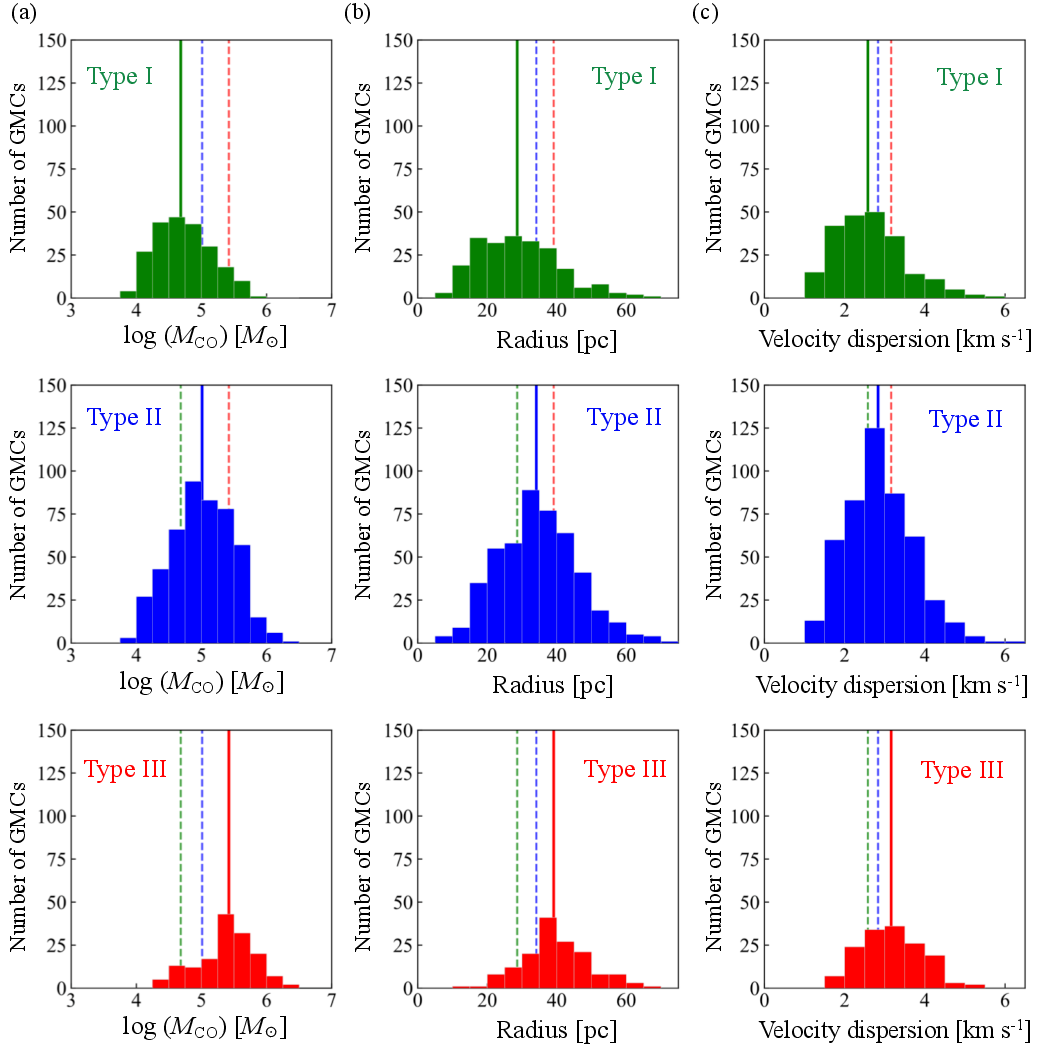}
 \end{center}
 \caption{Histograms of (a) $M_{\rm CO}$, (b) radius, and (c) velocity dispersion of GMCs in each type. 
The upper, middle, and lower panels indicate the properties of Type I, II, and III GMCs, respectively. 
Green, blue, and red lines indicate the median values of the histogram for each of Type I, II, and III GMCs, respectively.
}\label{fig:GMCprop}
\end{figure*}

Figure~\ref{fig:GMCprop} shows the frequency distribution of $M_{\rm CO}$, radius, and velocity dispersion for the three types of GMCs.
The vertical solid lines represent the median values of the histograms.
We found that the mass, size, and velocity dispersion systematically increase in the order of Type I, II, and III GMCs. 
Focusing on the mass distribution, typical masses of Type II and Type III GMCs are larger than 10$^{5}$\,$M_{\odot}$ whereas that of Type I GMCs is smaller than $10^{5}$\,$M_{\odot}$. 
This suggests that high-mass star-forming GMCs likely have their mass of $>$ 10$^{5}$\,$M_{\odot}$.
However, a significant number of star-forming GMCs exhibit masses lower than 10$^{5}$\,$M_{\odot}$, comprising nearly half of the Type II GMCs.
We consider that such low-mass Type II and Type III GMCs are now dissipating after high-mass star formation due to photo ionization and/or dissociation by UV radiation from high-mass stars (e.g., \cite{Inutsuka15}).
Alternatively, low-mass GMCs are probably evolving into more massive GMCs up to $M_{\rm CO}$ of about 10$^{6}$\,$M_{\odot}$ and undergoing a transition to more active star formation (see the detailed discussion presented in Section~\ref{D:spiralformation}).

There are a small number of massive ($>$ 10$^{5.5}$\,$M_{\odot}$) Type I GMCs.
Such inactive star-forming GMCs are highly rare in the Milky Way except for a few examples (\cite{Maddalena85}, Dobashi et al.\,\yearcite{Dobashi94},\yearcite{Dobashi96}).
Besides, in M33, GMC-8 with a total molecular gas mass of $\sim10^{6}$\,$M_{\odot}$ is also an exclusive sample \citep{Kondo21}.
Although GMC-8 is just associated with small H$\alpha$ emissions with the total \textit{L}\,(H$\alpha$) of $\sim$10$^{36.6}$\,erg\,s$^{-1}$ and is classified into Type II GMCs in this work, its star formation is relatively inactive considering the rich molecular reservoir (see also Section~\ref{R:indivregions}).
These GMCs are considered to be at a very young stage; they are formed recently in a short time of a few Myr \citep{Maddalena85,Kondo21}, which may evolve into the high-mass star-forming phase.

Note that we need to consider the variations of $R_{21}$ and $X_{\rm CO}$ when we calculate the GMC mass using the $^{12}$CO($J$=2--1) luminosity.
$R_{21}$ varies from region to region within an individual galaxy, and $R_{21}$ has a positive correlation with the star-formation rate (e.g., \cite{yajima2021}).
Also in M33, $R_{21}$ varies in each region and its error of $\sim$\,30\,$\%$ when considering the assumption of a constant $R_{21}$ of 0.6 over the M33 disk \citep{Muraoka23}. 
If we assume that such an error of 30\,$\%$ in $R_{21}$ dominates the uncertainty in the GMC mass,
this yields the mass medians of (4.8\,$\pm$\,1.4)\,$\times\,$10$^{4}$, (1.0\,$\pm$\,0.3)\,$\times\,$10$^{5}$, and (2.6\,$\pm$\,0.8)\,$\times\,$10$^{5}$ for Type I, II, and III GMCs, respectively.
Thus, the increasing trend in GMC mass from Type I to Type III remains unchanged.
Besides, $X_{\rm CO}$ changes depending on environmental parameters such as gas surface density and excitation state, and there is a factor of 1.3 uncertainty in the $X_{\rm CO}$ for the inner disk of the Milky Way \citep{Bolatto13}. 
\citet{Wall16} also reported that $X_{\rm CO}$ in the inter-arm regions is higher than in the arms by factors of 1.5 -- 2 for M51 and M83.
However, the spiral structures are less prominent in M33 than in grand-design spiral galaxies such as M51 and M83,
thus we expect the difference in $X_{\rm CO}$ between arm and inter-arm of M33 to be smaller.

\begin{table*}[htbp]
\tbl{GMC type and properties of the GMCs. }{%
\begin{tabular}{@{}l@{\hspace{15.5mm}}c@{\hspace{15.5mm}}c@{\hspace{15.5mm}}c@{}}  
\noalign{\vskip3pt} \hline \hline
\noalign{\vskip2pt}
GMC Type & Type I & Type II & Type III \\
\noalign{\vskip2pt}
\hline
\noalign{\vskip2pt}
Number  & 224 (26\,$\%$)  & 473 (56\,$\%$) & 151 (18\,$\%$)\\
Mass range ($M_{\odot}$)  & 6.7$\times\,$10$^{3}$ -- 6.0$\times\,$10$^{5}$  & 8.7$\times\,$10$^{3}$ -- 2.6$\times\,$10$^{6}$ & 1.9$\times\,$10$^{4}$ -- 2.1$\times\,$10$^{6}$ \\
Mass median ($M_{\odot}$)  & 4.8$\times\,$10$^{4}$  & 1.0$\times\,$10$^{5}$ & 2.6$\times\,$10$^{5}$ \\
Radius range (pc)  & 7 -- 67 & 8 -- 72 & 10 -- 68 \\
Radius median (pc)  & 29 & 34 & 39 \\
Velocity dispersion range (km\,s$^{-1}$) & 1.0 -- 5.8 & 1.1 -- 6.1 & 1.8 -- 5.4 \\
Velocity dispersion median (km\,s$^{-1}$) & 2.6 & 2.8 & 3.2 \\
$M_{\rm Vir}/M_{\rm CO}$ median & 3.1 & 2.3 & 1.5 \\
$^{13}$CO detection rate & 8/224 $\sim$4\,$\%$ & 95/473 $\sim$20\,$\%$ & 76/151 $\sim$50\,$\%$ \\
\textit{R}$_{13/12}$ median & -- & 0.11 & 0.12 \\
\noalign{\vskip2pt}
\hline
\end{tabular}}
\label{table:clsum}
\end{table*}

\begin{figure*}[htbp]
 \begin{center}
  \includegraphics[width=160mm]{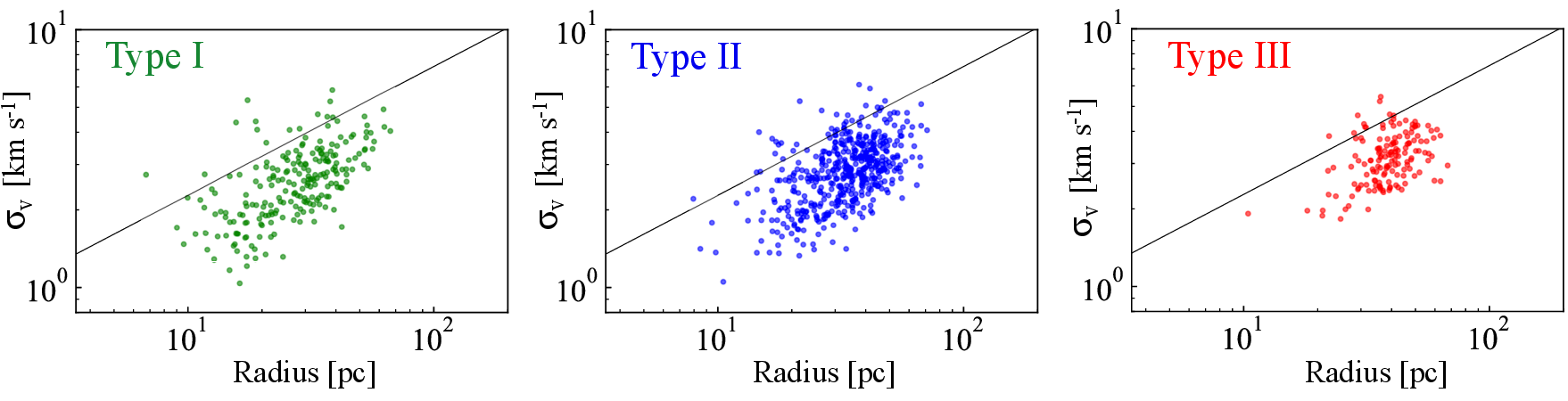}
 \end{center}
 \caption{Radius-velocity dispersion (\textit{R} - $\sigma_v$) relation for each type of GMC. 
Green, blue, and red dots indicate Type I, II, and III GMCs, respectively.
The solid line represents the relation for the Milky Way clouds ($\sigma_v$ = 0.72\,$R^{0.5}$: \cite{solomon1987}).
}\label{fig:sizevsigma}
\end{figure*}

\subsection{Size-Line-width Relation}\label{R:sizelinewidth}
Galactic molecular clouds show correlations between radius $R$ in pc and velocity dispersion $\sigma_v$ in km\,s$^{-1}$, which is expressed as $\sigma_v$ = 0.72\,$R^{0.5}$ \citep{solomon1987}.
On a GMC scale, \citet{Hirota11} discovered that the GMCs with H$\;${\sc ii} regions located in the downstream of the spiral arm in IC~342 have smaller $\sigma_v$ compared to the GMCs without H$\;${\sc ii} regions in the upstream. They suggest that turbulence in the GMCs dissipates and the GMCs become more gravitationally bound by the effects of the arm, resulting in the high-mass star formation.
Contrary to this, on sub-pc scale, \citet{Wong19} found that the molecular cloud in the brightest star-forming region 30~Doradus in the LMC has a larger $\sigma_v$ than the quiescent cloud with very low star formation activity.
The higher $\sigma_v$ in 30~Doradus is considered to be due to gravitational collapse or energetic stellar feedback.
These two studies focus on individual GMCs but show different trends of the \textit{R} - $\sigma_v$ relation. 
The \textit{R} - $\sigma_v$ relation in molecular clouds is still being studied as a function of star formation activity in different environments and at different physical scales, which motivates us to investigate its fundamental properties in our M33 study as well.

In Figure~\ref{fig:sizevsigma}, we compare the \textit{R} - $\sigma_v$ relation between the three types of GMCs across the entire molecular gas disk of M33.
The relation for the Milky Way clouds ($\sigma_v$ = 0.72\,$R^{0.5}$: \cite{solomon1987}) is shown as a line.
We cannot find significant differences in velocity dispersion at a given radius among the three types of GMCs. 
However, the distribution patterns on the plot vary according to the GMC types.
While the data points of Type I GMCs are spread out across the plot, those of Type III GMCs exhibit a clustered distribution.
Type II GMCs seem to exhibit a distribution that falls between Type I and Type III GMCs.
In other words, quiescent GMCs show significant variability in both radius and velocity dispersion but active star-forming GMCs display show similar properties than quiescent GMCs. 
A plausible interpretation of this trend is that there is physical thresholds for high-mass star formation within GMCs, and Type I GMCs encompass both GMCs which are likely to undergo star formation in the near future and those that continue to drift in interstellar space without high-mass star formation.

\subsection{CO Luminosity Mass–Virial Mass Relation}\label{R:McoMvir}
\begin{figure}[htbp]
 \begin{center}
  \includegraphics[width=80mm]{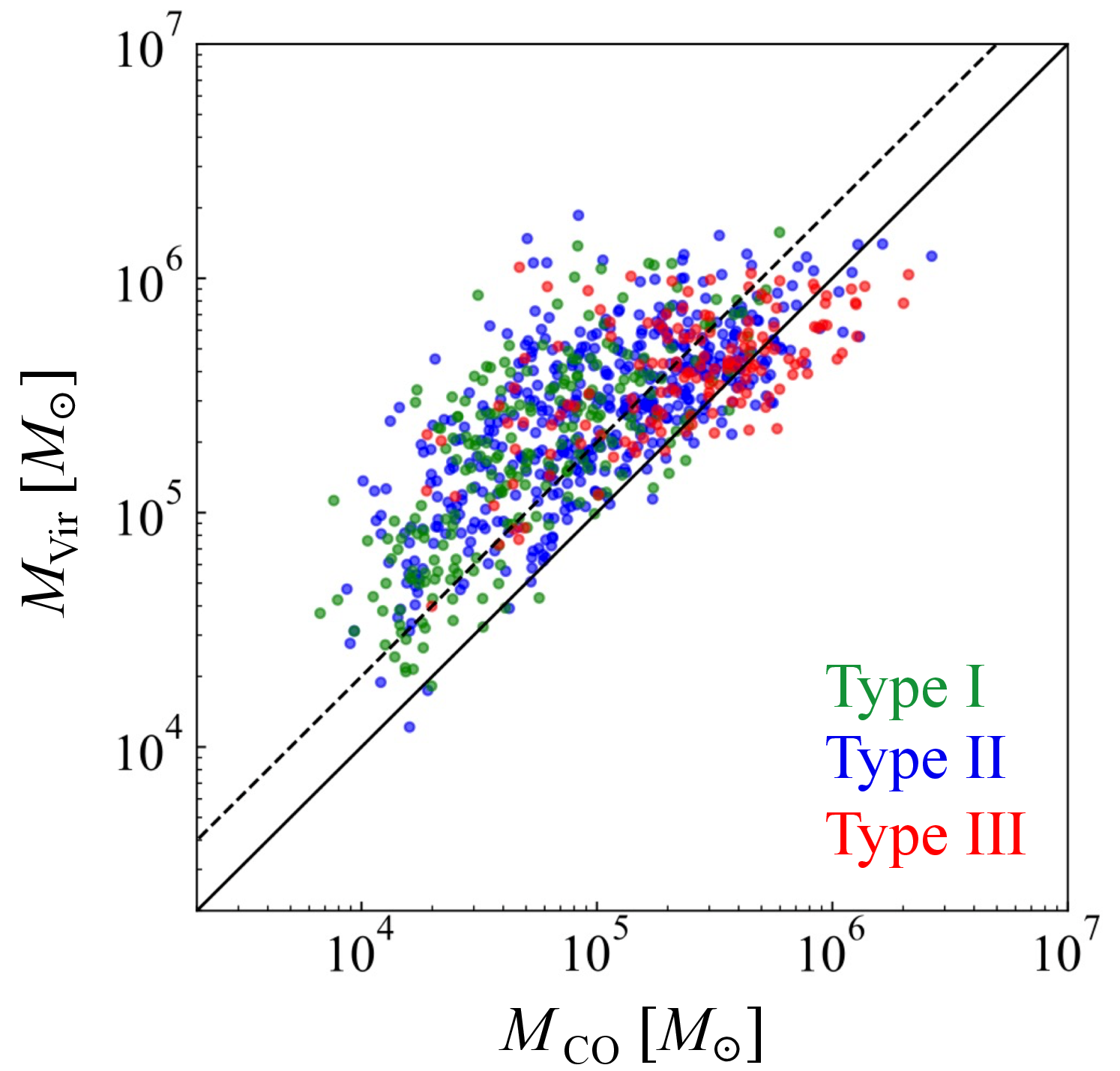}
 \end{center}
 \caption{$M_{\rm CO}$ and $M_{\rm Vir}$ relation for each GMC type. Green, blue, and red dots indicate Type I, II, and III GMCs, respectively. The solid and dashed line represents $M_{\rm Vir}$ = $M_{\rm CO}$ and $M_{\rm Vir}$ = 2\,$M_{\rm CO}$, respectively.
}\label{fig:McoMvir}
\end{figure}

The virial parameter, defined as the ratio of the $M_{\rm CO}$ to the virial mass ($M_{\rm Vir}$), is considered to reflect the ratio of the turbulent kinetic energy to the self-gravitational energy of GMCs.
The virial mass is calculated as
\begin{eqnarray}
M_{\rm Vir} = 1040\,R\,\sigma_v^2
\end{eqnarray}
where $R$ is the radius in parsec, and $\sigma_v$ is the velocity dispersion in km\,s$^{-1}$ \citep{solomon1987}. 
Figure~\ref{fig:McoMvir} shows the relation between $M_{\rm CO}$ and $M_{\rm Vir}$ for three types of GMCs. 
$M_{\rm Vir}$ and $M_{\rm CO}$ are well correlated, but $M_{\rm Vir}$ is generally larger than $M_{\rm CO}$.
The detail of this trend is discussed in \citet{Muraoka23}.
The median values of the virial parameter for Type I, II, and III GMCs are 3.1, 2.3, and 1.5, respectively. 
This suggests that active star-forming GMCs are closer to virial equilibrium than the quiescent GMCs.

\begin{figure*}[bthp]
 \begin{center}
  \includegraphics[width=160mm]{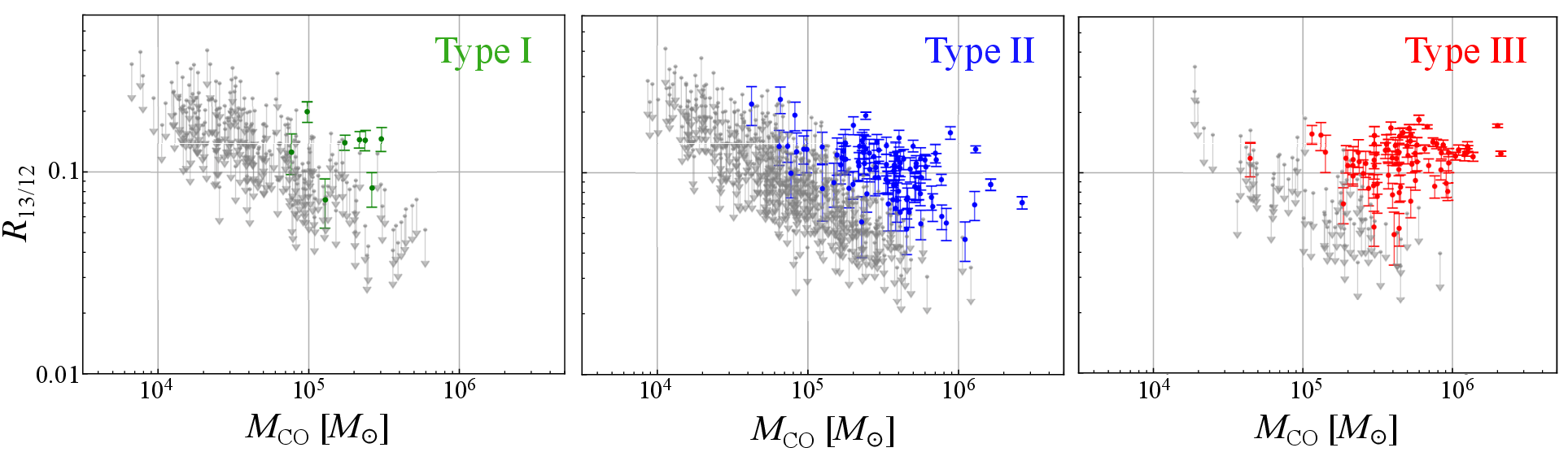}
 \caption{Scatter plots of $M_{\rm CO}$ and \textit{R}$_{13/12}$ for each GMC type. The gray circles and arrows indicate the 3\,$\sigma$ upper limit of \textit{R}$_{13/12}$ for $^{13}$CO non-detected GMCs.
}\label{fig:1312ul}
\end{center}
\end{figure*}

\subsection{Dense gas tracer $^{13}$CO emission}\label{R:13CO}
Star formation activities are closely related to denser gas regions of molecular clouds (e.g., Gao \& Solomon\,\yearcite{Gao04a}, \yearcite{Gao04b}; Muraoka et al.\,\yearcite{Muraoka09a}, \yearcite{Muraoka09b}; \cite{Lada10}; Chen et al.\,\yearcite{Chen15}, \yearcite{Chen17}; \cite{Shimajiri17, Querejeta19, Torii19}).
$^{13}$CO lines are mainly detected in dense region with a H$_2$ volume density of $\gtrsim$\,10$^{3}$\,cm$^{-3}$ within the $^{12}$CO cloud boundaries (e.g., \cite{Nishimura15}).
In M33, a large number of $^{13}$CO($J$=2--1) sources are detected over the molecular-gas disk \citep{Muraoka23}.
In this section, we examined the correlation of the $^{13}$CO properties with each type of GMC.

Firstly, we compared the detection rate of $^{13}$CO emission between three types of GMCs (Table~\ref{table:13COdetection}). The criteria for determining the detection are that $^{13}$CO emission exceeds 4\,$\sigma$ in continuous over two velocity channels or exceeds 3\,$\sigma$ in continuous over three velocity channels.
We detected significant $^{13}$CO emission for 8 Type I GMCs, 95 Type II GMCs, and 76 Type III GMCs, yielding the detection rate of 4\,$\%$, 20\,$\%$, and 50\,$\%$.
The $^{13}$CO detection rate clearly increases from Type I to Type III GMCs.

\begin{table}[htbp]
\tbl{The $^{13}$CO($J$=2--1) detection rate for each GMC type.\footnotemark[$*$] }{%
\begin{tabular}{@{}l@{\hspace{-5.5mm}}ccc}  
\noalign{\vskip3pt} \hline \hline
\parbox[c][20pt][c]{100pt}{Mass range ($M_{\odot}$)}  &\, Type I \,& \,Type II \,& \,Type III \,\\
\noalign{\vskip3pt}
 \hline
 \noalign{\vskip2pt}
\parbox[c][20pt][c]{80pt}{All} & \(\displaystyle \frac{8}{224}\) $\sim$4\,$\%$  & \(\displaystyle \frac{95}{473}\) $\sim$20\,$\%$ & \(\displaystyle \frac{76}{151}\) $\sim$50\,$\%$ \\
\parbox[c][20pt][c]{80pt}{low ($M_{\rm CO}$ $<$ 10$^{5.0}$)} & \(\displaystyle \frac{2}{165}\) $\sim$1\,$\%$ & \(\displaystyle \frac{9}{223}\) $\sim$4\,$\%$ & \(\displaystyle \frac{1}{30}\) $\sim$3\,$\%$ \\
\parbox[c][20pt][c]{80pt}{high ($M_{\rm CO}$ $\geqq$ 10$^{5.0}$)}   & \(\displaystyle \frac{6}{59}\) $\sim$5\,$\%$ & \(\displaystyle \frac{86}{240}\) $\sim$35\,$\%$ & \(\displaystyle \frac{75}{121}\) $\sim$62\,$\%$ \\
\noalign{\vskip2pt}
\hline 
\end{tabular}}
\begin{tabnote}\label{table:13COdetection}
{\hbox to 0pt{\parbox{78mm}{\footnotesize
\par\noindent
\footnotemark[$*$] Detection rate $=$ (Number of $^{13}$CO-detected GMCs)  / (Total number of GMCs)
\par\noindent}}}
\end{tabnote}
\end{table}

However, the spectral intensity of $^{13}$CO emission is sensitive to the column densities (and thus the mass) of molecular clouds.
$M_{\rm CO}$ increases from Type I to Type III GMCs as presented in Section~\ref{R:bprop}, which may result in the mass-dependency of the $^{13}$CO detection.
To make a fair comparison between the three types of GMCs, we derived the $^{13}$CO detection rate by dividing the GMCs into two subsamples; low-mass ($M_{\rm CO}$\,$<$\,10$^{5.0}$\,$M_{\odot}$) and high-mass ($M_{\rm CO}$\,$\geqq$\,10$^{5.0}$\,$M_{\odot}$) GMCs (Table~\ref{table:13COdetection}).
While there are little difference between three types in the low-mass range, the $^{13}$CO detection rates increase in the order of Type I, II, and III GMCs in the high-mass range.
Such a remarkable difference in $^{13}$CO detection rate for each type suggests that star formation occurs in dense gas, and there is a strong correlation between dense gas formation and active star formation (e.g., \cite{Tokuda23}). Note that the detection rates are considered to be strictly limited by the sensitivity of our $^{13}$CO data because any type of GMC has a very low detection rate in the low-mass range.

Figure~\ref{fig:1312ul} shows the relation between $M_{\rm CO}$ and $^{13}$CO/$^{12}$CO integrated intensity ratio \textit{R}$_{13/12}$. We estimated \textit{R}$_{13/12}$ at the peak position of $^{12}$CO intensity $W\rm (^{12}CO)$ in the GMCs using the $^{13}$CO($J$=2--1) data obtained by ACA 7\,m array. 
Gray circles indicate the 3\,$\sigma$ upper limit of \textit{R}$_{13/12}$ for $^{13}$CO non-detected GMCs,
which are defined as follows:
\begin{equation}\label{eq:1312upperlimit}
 3\,\sigma\,(upper\,limit) = \frac{3\,\times\,W(\rm \sigma_{\rm ^{13}CO})}{W(\rm ^{12}CO)},
\end{equation}
\begin{equation}
W(\rm \sigma_{\rm ^{13}CO}) = \sigma_{\rm ^{13}CO} \times \Delta \textit{V}_{\rm chan} \times (\textit{N}_{\rm chan})^{0.5}
\end{equation}
where $W(\rm \sigma_{\rm ^{13}CO})$ is the rms of the $^{13}$CO integrated intensity, $\sigma_{\rm ^{13}CO}$ is the rms noise level of the $^{13}$CO spectrum, $\Delta V_{\rm chan}$ is the velocity channel width of 1.4\,km\,s$^{-1}$, and $N_{\rm chan}$ is the number of velocity channels for $^{13}$CO data, which are estimated at the $^{12}$CO intensity peak of $^{13}$CO non-detected GMCs.   
In cases where the upper limit is high, there are little constraints on the true \textit{R}$_{13/12}$.
The upper limits exhibit a clear trend that they moves downward with the increase in $M_{\rm CO}$.
This is because while $W(\rm ^{12}CO)$ increases, $W(\rm \sigma_{\rm ^{13}CO})$ remains unchanged.
Therefore, it is difficult to obtain the accurate $^{13}$CO detection rate at the low-mass range ($M_{\rm CO}$ $<$ 10$^{5.0}$\,$M_{\odot}$, see Table~\ref{table:13COdetection}), especially for Type I GMCs with a typical mass of lower than 10$^{5.0}\,M_{\odot}$.

\begin{figure}[htbp]
 \begin{center}
  \includegraphics[width=65mm]{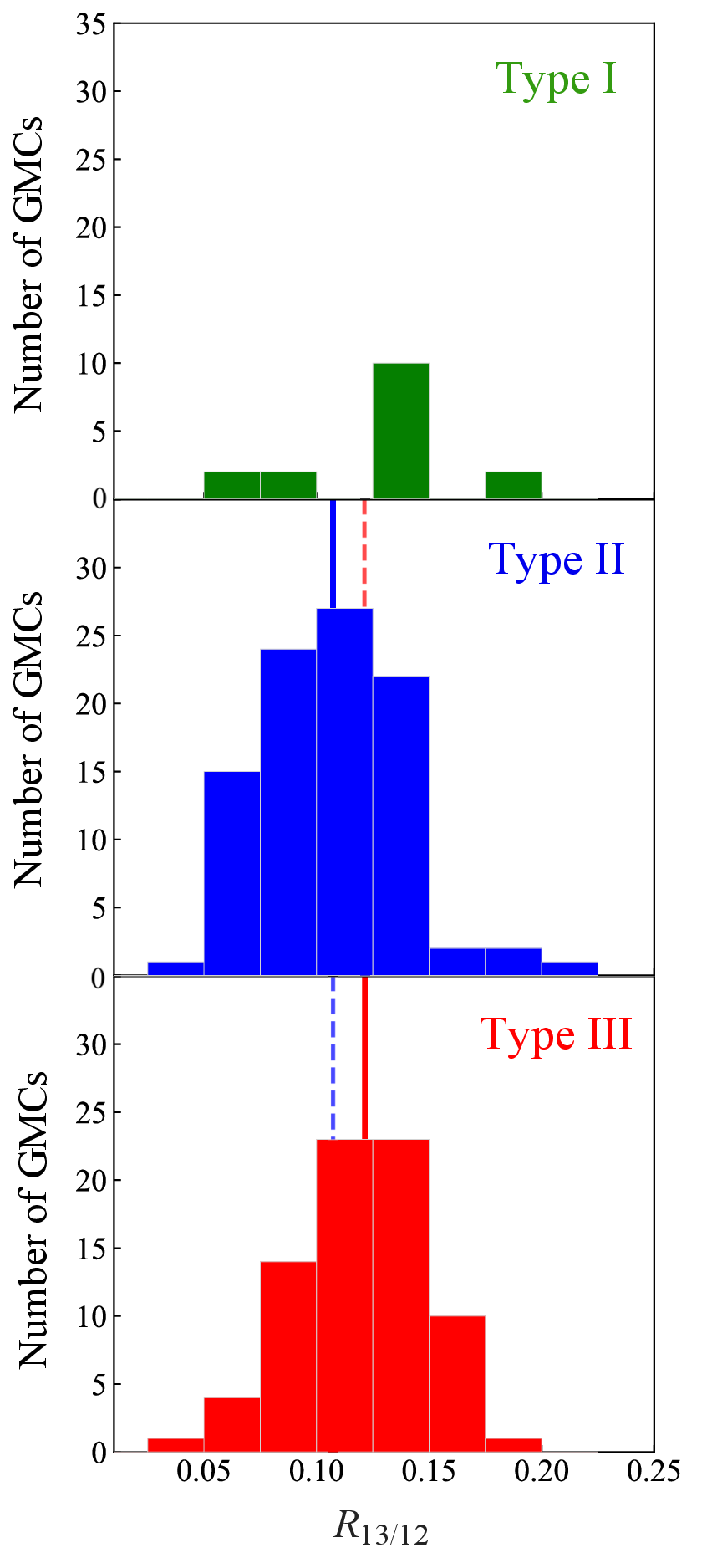}
 \end{center}
 \caption{Histogram of \textit{R}$_{13/12}$ for each type of GMC. 
Blue and red lines indicate the median for histograms of Type II and Type III GMCs, respectively.
}\label{fig:1312ratio}
\end{figure}

Figure~\ref{fig:1312ratio} shows the \textit{R}$_{13/12}$ histograms for $^{13}$CO-detected GMCs. 
The \textit{R}$_{13/12}$ median of Type II and Type III GMCs are 0.11 and 0.12, respectively, where that of Type III GMCs is slightly larger.
In order to test whether the \textit{R}$_{13/12}$ frequency distributions are different, we performed a two-sample Kolmogorov–Smirnov test and obtained a p-value of 0.012.
This means that we can reject the null hypothesis that \textit{R}$_{13/12}$ frequency distributions of Type II and Type III GMCs originate from the same distribution with a significance level of 5$\%$.
These results suggest that the dense gas fraction in GMCs increases as the high-mass star formation progresses (see also \cite{Kondo21}), which is consistent with the $^{13}$CO detection rate of each type of GMC.

\begin{figure*}[htbp]
 \begin{center}
  \includegraphics[width=160mm]{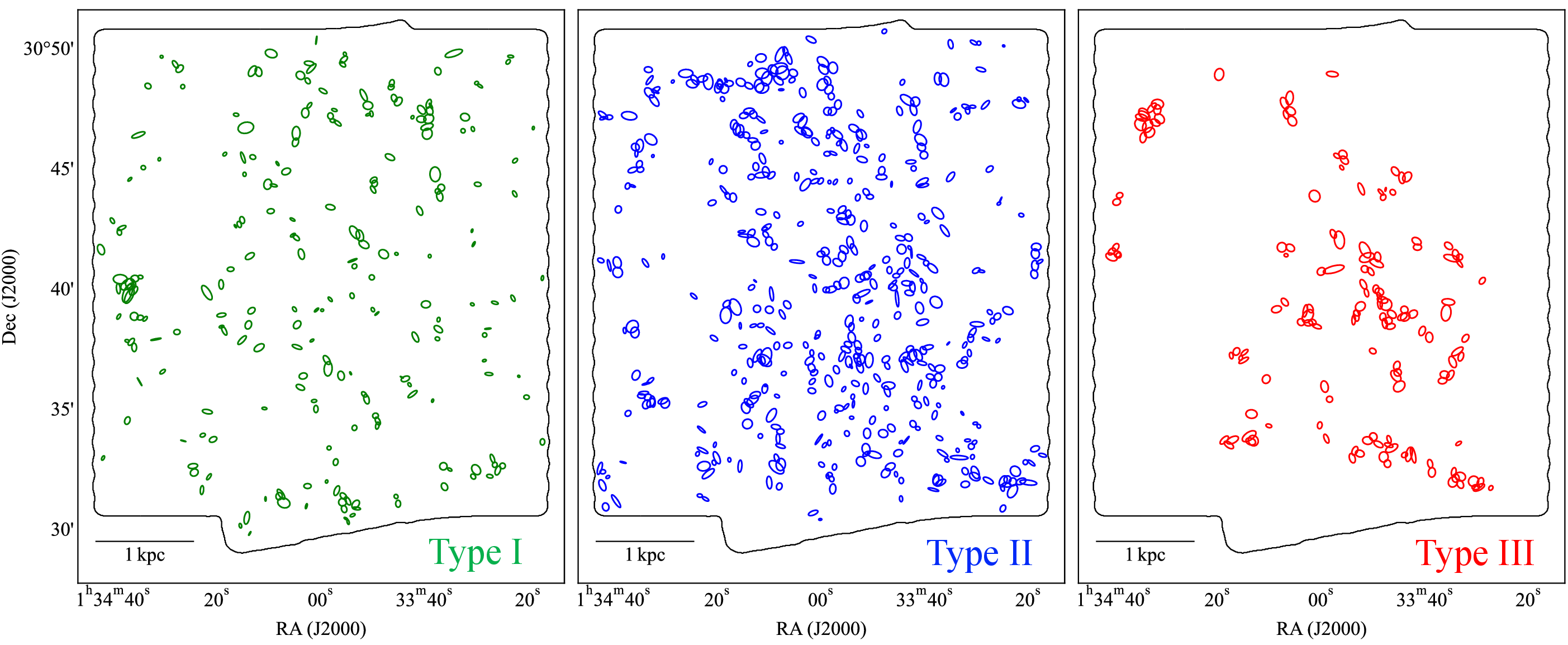}
 \end{center}
 \caption{Spatial distributions of three types of GMCs in the M33 disk. Green, blue, and red ellipses indicate Type I, II, and III GMCs, respectively. The enclosure by the black line shows the ACA observation field. We can see the remarkable differences in the randomness of the GMC distributions according to their types.
}\label{fig:onlytypeGMCs}
\end{figure*}

\begin{figure}[htbp]
 \begin{center}
  \includegraphics[width=70mm]{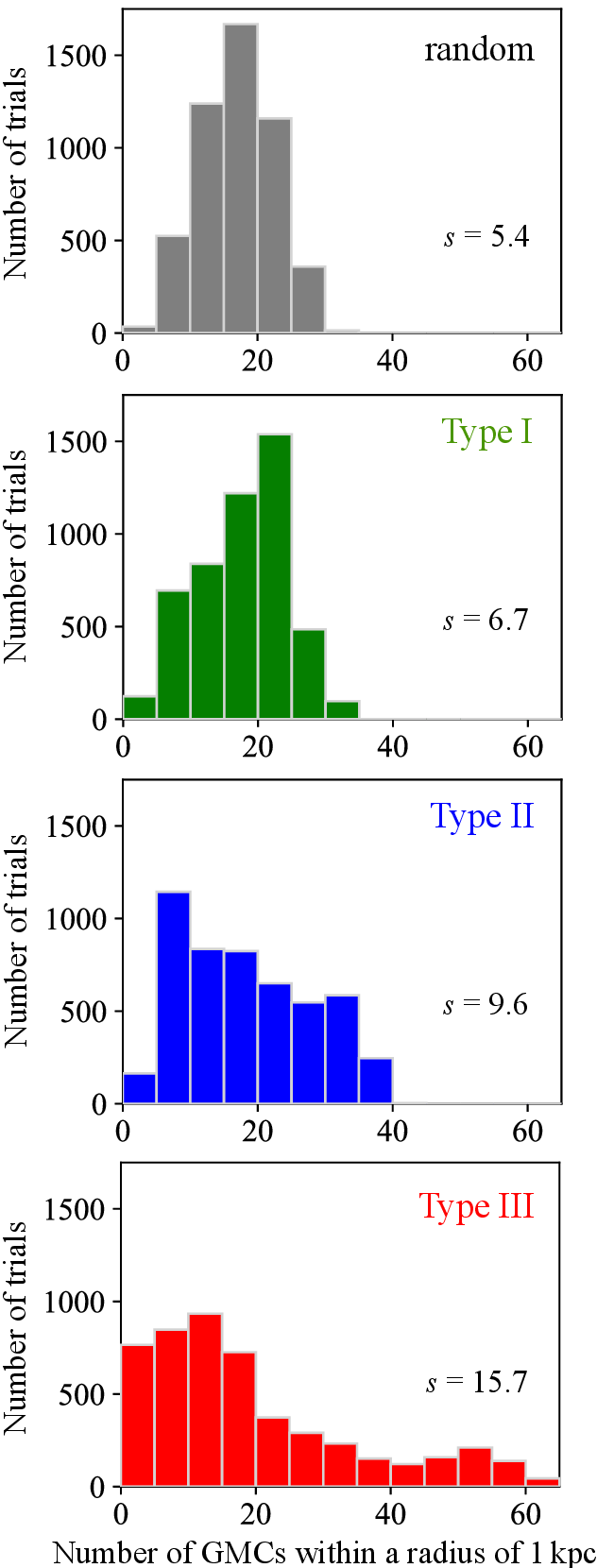}
 \end{center}
 \caption{Histogram of the number of GMCs within a circular aperture with a radius of 1\,kpc which was randomly located in the ACA observation field. The horizontal axes for Type I and Type II GMCs are normalized to the number of Type III GMCs. Gray one indicates the expected distributions if the same number of Type III GMCs are distributed at random.
}\label{fig:numcount}
\end{figure}

\subsection{Spatial distribution of GMCs}\label{R:spiral}
Figure~\ref{fig:onlytypeGMCs} shows the spatial distribution of the three types of GMCs in the M33 disk.
We found remarkable differences in the randomness of the GMC distributions according to their types.
Many Type I GMCs are distributed randomly and sparsely across the disk.
The randomness of the Type II GMC distribution becomes weaker than that of Type I GMCs; Type II GMCs seem gathering together. 
The distribution of Type III GMCs is spatially localized, and exhibits spiral structures.

To quantitatively evaluate these spatial randomness of the three types of GMCs, 
we counted the number of each type of GMCs within a circular aperture with a radius of 1\,kpc which was randomly located in the ACA observation field.
Then this counting process is repeated 5000 times, and the counted number of GMCs, which is normalized to the number of Type III GMCs, is plotted as histograms (Figure~\ref{fig:numcount}). 
The gray histogram illustrates the expected distributions assuming the same number of Type III GMCs are randomly distributed.
We performed 100 iterations to generate these random distributions and subsequently computed their mean distribution, with the standard deviation \textit{s} of 5.4.
The distribution of Type I GMCs are similar to the random distribution (\textit{s} = 6.7), and that of Type II GMCs are flatter than Type I GMCs (\textit{s} = 9.6).
The histogram of Type III GMCs is widespread and has the highest \textit{s} of 15.7, which indicates that the number density of Type III GMCs significantly varies from region to region.

These results suggest that GMCs can be formed anywhere in M33, while the star formation activities are strongly influenced by the galactic environment, especially by galactic dynamics around spiral arms.
A detailed discussion of the relationship between GMCs and spiral arms is given in Section~\ref{D:spiralformation}.

\section{Young clusters associated with GMCs}\label{R:ysc}
Clusters are born within molecular clouds, and clusters in the earliest stages are embedded in dense dust, which is traced by mid- and/or far-infrared radiation.
Then clusters are blowing the surrounding materials and become optically visible as they evolve in time.
In this section, we attempt to confirm the validity of the H$\alpha$-luminosity-based classification of GMCs by comparing the spatial distribution and properties between GMCs and YSCs.

\subsection{Association between GMCs and clusters}\label{R:asysc}
We used two cluster catalogues from \citet{Meulenaer15} and \citet{Corbelli17}, whose cluster-age ranges are 10$^{6.6}$ -- 10$^{10.1}$\,yr and 10$^{6.0}$ -- 10$^{7.1}$\,yr, respectively. 
Note that we remove YSC candidates, whose ages and masses have not been determined by \citet{Sharma11}, from the \citet{Corbelli17} catalogue.
\citet{Corbelli17} young clusters (C-YSCs) and \citet{Meulenaer15} clusters (M-SCs) differ in the methods and wavelengths used to derive the cluster parameters (as described in Section~\ref{da:sc} and the Appendix~\ref{app:clct}). 

\begin{figure}[htbp]
 \begin{center}
  \includegraphics[width=80mm]{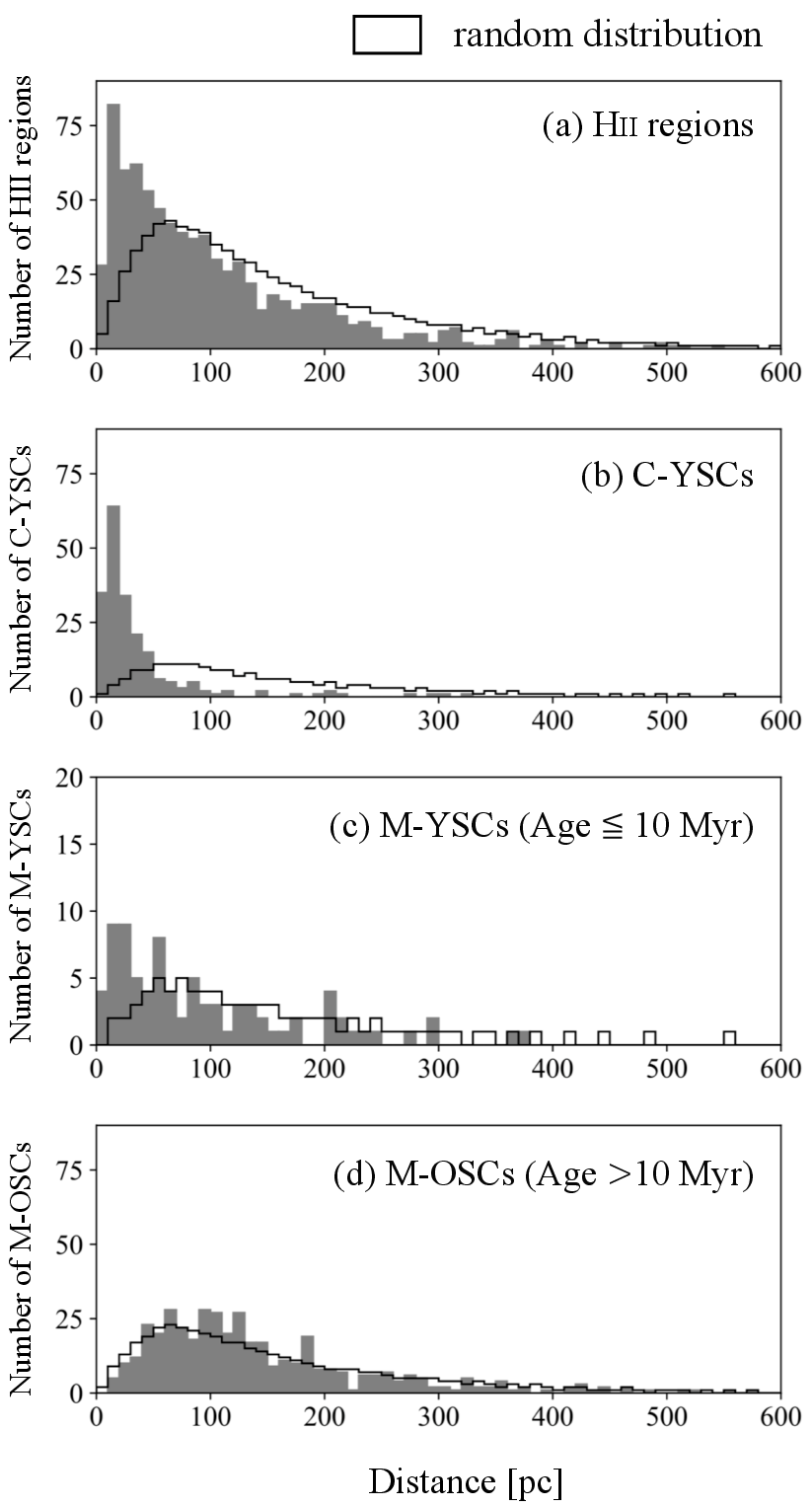}
 \end{center}
 \caption{Frequency distributions of the deprojected distance of (a) H$\,${\sc ii} regions, (b) C-YSCs, (c) M-YSCs, and (d) M-OSCs from the nearest GMC in the ACA observation field. 
Open white columns show the distribution expected if the same number of H$\,${\sc ii} regions and clusters are distributed at random in the observed area. 
}\label{fig:coyscdistance}
\end{figure}

Figure~\ref{fig:coyscdistance} shows the frequency distributions of the deprojected distance of H$\,${\sc ii} regions and clusters from the nearest GMCs.
We divided M-SCs into two groups: one consists of clusters younger than 10\,Myr (M-YSCs), and the other consists of clusters older than 10\,Myr (M-OSCs).
The open white columns represent the distributions expected if the same number of the H$\,${\sc ii} regions and clusters are distributed at random in the ACA observation field. 
We generated these random distributions 100 times and calculated the mean distribution.
These frequency distributions indicate that the H$\,${\sc ii} regions and C-YSCs are spatially better correlated with GMCs than the random distribution, and also than M-YSCs and M-OSCs.  
The histogram shape of M-YSCs is almost flat and M-YSCs seem to have a weaker correlation than C-YSCs by eye.
The reason for such a difference in the spatial correlation with GMCs between C-YSCs and M-YSCs is because M-YSCs based on optical observations are considered to be older than C-YSCs identified using far-infrared 24\,$\mu$m data.
C-YSCs include younger clusters at the embedded star formation phase than M-YSCs (see also Figure~\ref{fig:ysccatalog} in the Appendix~\ref{app:clct}), which results in a strong correlation with GMCs. 
In addition, we consider that the small sample size for M-YSCs results in its weak correlation with GMCs.
Hereafter, we focus on only C-YSCs for comparison with GMCs.

It is consistent with the previous studies (e.g., \cite{Kawamura09}, \cite{Peltonen23}, \cite{Demachi23}) that younger clusters are spatially better correlated with GMCs than older clusters. 
Also, the frequency distribution of C-YSCs shows a sharper peak than that of H$\,${\sc ii} regions.
We consider that H$\,${\sc ii} regions distant from GMCs are associated with molecular clouds below detection limit or used to have GMCs but they have already dissipated due to stellar feedback.
However, in the LMC \citep{Kawamura09}, the frequency distribution of the deprojected distance of H$\,${\sc ii} regions from the nearest GMC shows a peak as sharp as that of young clusters.
This difference is considered to be mainly due to the spatial resolution of each H$\alpha$ data.
H$\,${\sc ii} regions in the LMC are more spatially-resolved than those in M33, and thus some luminous H$\,${\sc ii} regions associated with GMCs (e.g., 30\,Doradus, N159, N44) are regarded as an aggregation of multiple H$\alpha$ sources (see Fig.9 in \cite{Kawamura09}), which increases the number of H$\,${\sc ii} regions near GMCs.

\begin{table*}[htbp]
\begin{center}
\tbl{Association between GMCs and young stellar clusters.\footnotemark[$*$] }{%
\begin{tabular}{@{}l@{\hspace{20mm}}c@{\hspace{18mm}}c@{\hspace{18mm}}c@{\hspace{4mm}}}  
\noalign{\vskip3pt} \hline \hline \noalign{\vskip2pt}
GMC type & Type I & Type II & Type III \\
\noalign{\vskip2pt} \hline
\noalign{\vskip2pt}
Total number of GMCs   & 224  & 473 & 151 \\
Number of GMCs w/ C-YSC(s) & 6 (3$\%$) & 140 (30$\%$) & 77 (51$\%$)\\
Number of GMCs w/o C-YSC(s) & 218 (97$\%$) & 333 (70$\%$) & 74 (49$\%$)\\
\noalign{\vskip2pt} \hline
\end{tabular}}
\begin{tabnote}
{\hbox to 0pt{\parbox{160mm}{\footnotesize
\par\noindent
\centering
\footnotemark[$*$] We defined the association of GMCs and C-YSCs if the position of a C-YSC is within the boundary of a GMC.
\par\noindent}}}
\end{tabnote}\label{table:YSCwGMC}
\end{center}
\end{table*}

Table~\ref{table:YSCwGMC} summarizes the association of C-YSCs and three types of GMCs.
The association was determined if the position of a C-YSC is within the boundary of a GMC. 
The total number of GMCs associated with C-YSCs is 223, and four GMCs of them with two C-YSCs.
Type III GMCs show the highest spatial correlation with C-YSCs, Type II GMCs moderate correlation, and Type I GMCs are almost uncorrelated.
We confirmed that all six Type I GMCs with C-YSCs have weak and compact H$\alpha$ emissions which are below the identification limit. 
Therefore, these Type I GMCs may actually be classified as Type II GMCs.
For most of Type III GMCs without C-YSCs, their spatial extents marginally overlap with H$\,${\sc ii} regions which have C-YSCs, but GMCs themselves do not overlap with the position of C-YSCs.
Some of these GMCs are part of GMC complexes such as e.g., NGC~604, NGC~595. 

\begin{figure*}[htbp]
 \begin{center}
  \includegraphics[width=160mm]{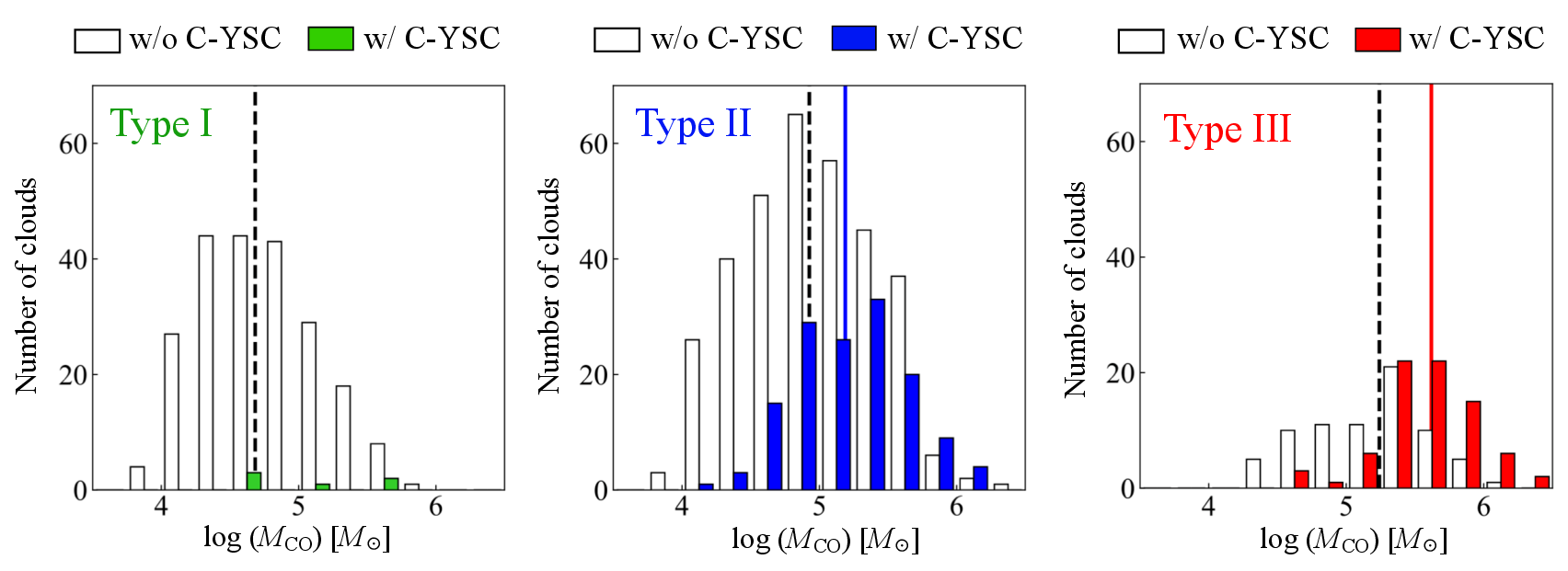}
 \end{center}
 \caption{The mass frequency distributions of Type I (left), Type II (center), and Type III (right) GMCs. 
Open white columns show the GMCs not associated with C-YSCs, and filled color (green, blue, red) columns the GMCs associated with C-YSCs.
Black dashed and colored solid lines indicate the median value for GMCs without C-YSCs and those with C-YSCs, respectively.
The median value of Type I GMCs with C-YSCs is not shown because of the small sample size.
}\label{fig:massyschist}
\end{figure*}

Figure~\ref{fig:massyschist} represents the mass distributions of three types of GMCs, which are associated with C-YSC(s) and not associated with C-YSC(s) shown by filled color (green, blue, red) columns and open white columns, respectively.
The median values of the mass for Type II and Type III GMCs are indicated by blue and red lines, respectively.
GMCs with C-YSCs are more massive than those without C-YSCs in both Type II and Type III GMCs.
The mass median of Type II and Type III GMCs associated with C-YSCs are $1.5\,\times\,10^{5}\,M_{\odot}$ and $4.2\,\times\,10^{5}\,M_{\odot}$, respectively.
One interpretation of these results is that star formation activity roughly depends on the amount of molecular gas.

\begin{figure*}[htbp]
 \begin{center}
  \includegraphics[width=155mm]{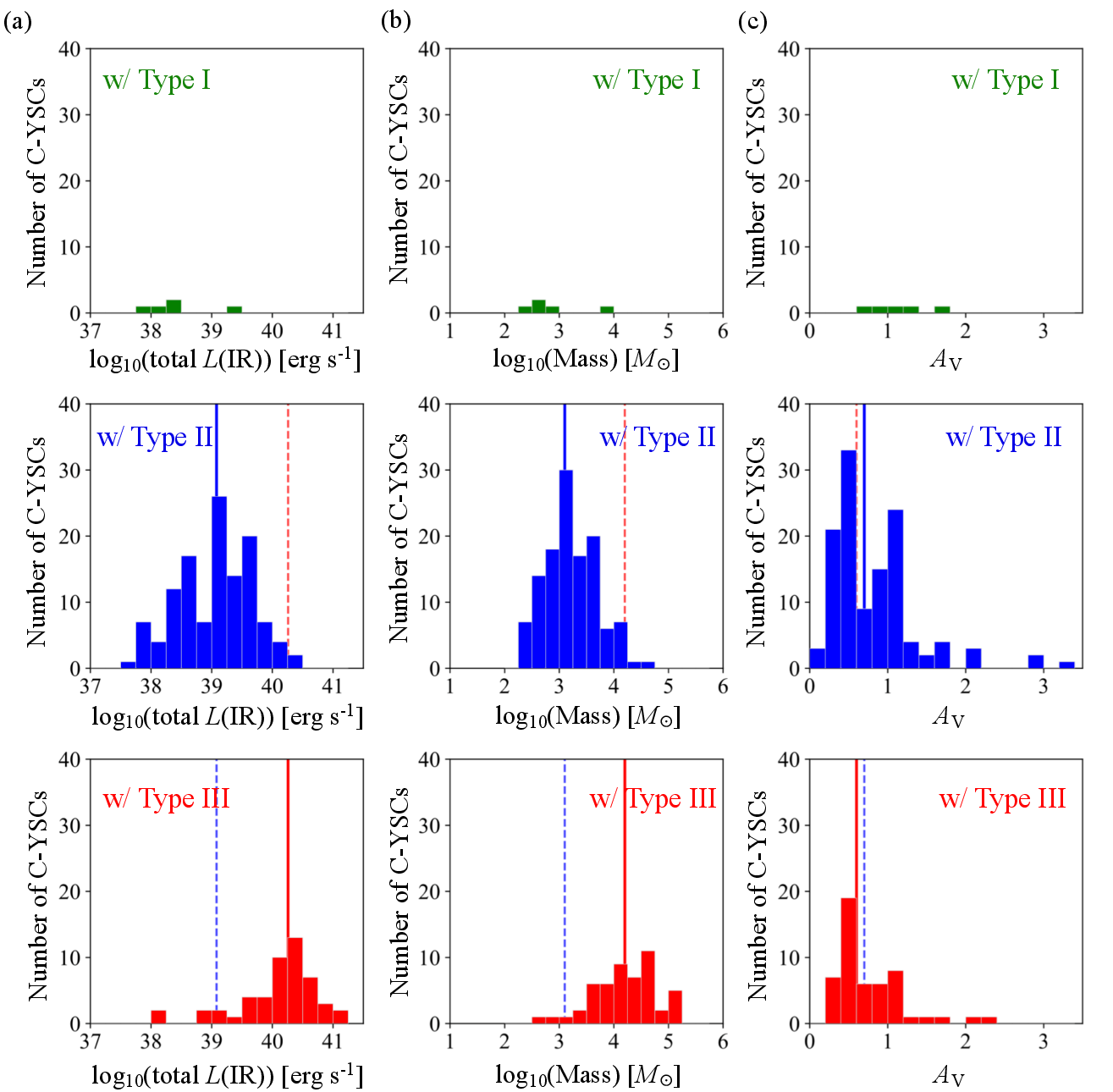}
 \end{center}
 \caption{Frequency distributions of C-YSC properties; (a) total infrared luminosity \textit{L}\,(IR), (b) mass, (c) visual extinction $A_{V}$. Green, blue, and red indicate the C-YSCs which are associated with Type I, II, and III GMCs, respectively. 
Blue and red lines indicate the median values of the properties of C-YSCs associated with Type II and Type III GMCs, respectively. 
}\label{fig:yscprop}
\end{figure*}

\subsection{Properties of clusters associated with the three types of GMCs}\label{R:yscprop}
We investigated the physical properties of C-YSCs associated with each type of GMC; the total infrared luminosity \textit{L}\,(IR), mass, and visual extinction $A_{V}$ (Figure~\ref{fig:yscprop}).
If a C-YSC is associated with two or more GMCs, we selected the nearest GMCs from the C-YSC.
The median values of the properties of C-YSCs associated with Type II and Type III GMCs are indicated by blue and red lines, respectively.
In Figures~\ref{fig:yscprop} (a) and (b), C-YSCs associated with Type III GMCs are more luminous in infrared and more massive than those with Type II GMCs, reflecting the differences in the star formation activity between the Type II and Type III GMCs. 
There are only five C-YSCs associated with Type I GMCs, thus they do not present any clear trend.
We found that the relation between $A_{V}$ and each GMC type is very weak, which indicates that dust extinction has little effect on the results shown in Figures~\ref{fig:yscprop} (a) and (b).

\begin{figure*}[htbp]
 \begin{center}
  \includegraphics[width=160mm]{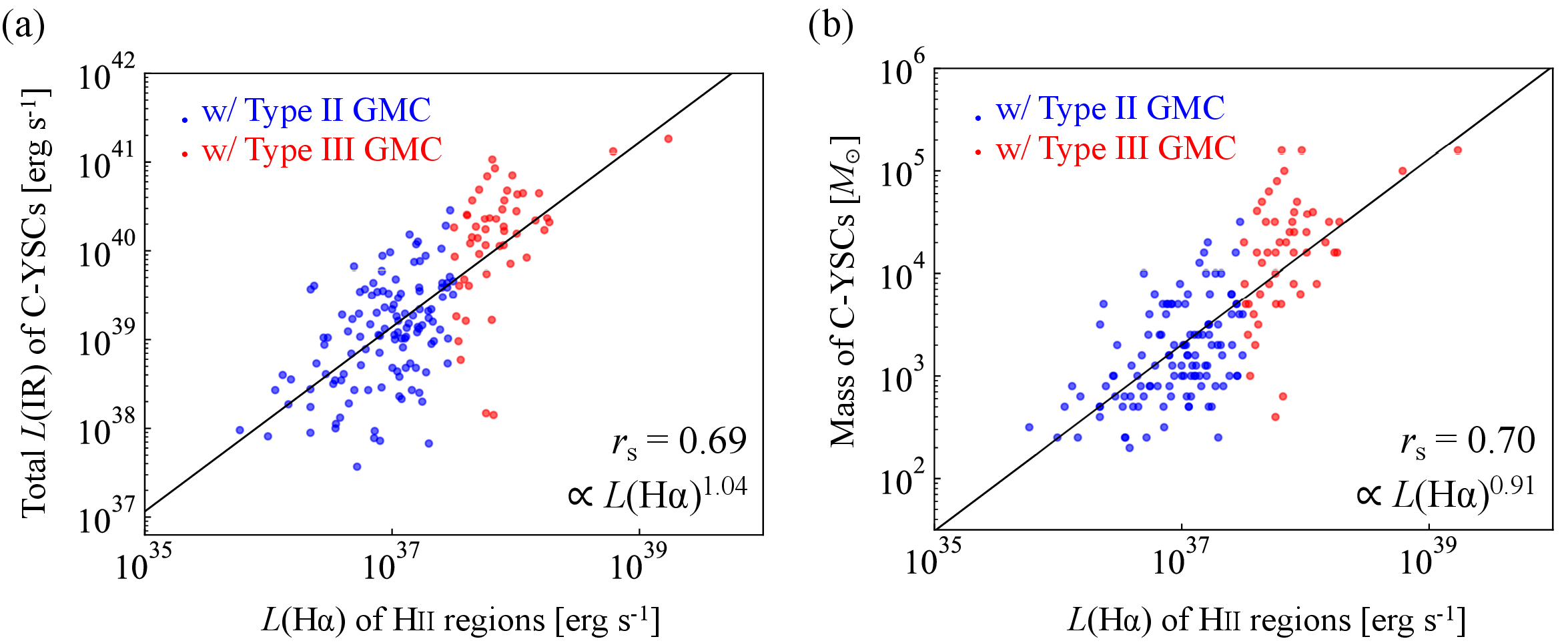}
 \end{center}
 \caption{Scatter plots comparing the \textit{L}\,(H$\alpha$) of the H$\,${\sc ii} regions and properties of C-YSCs; (a) total \textit{L}\,(IR), (b) mass.
Each H$\,${\sc ii} region and YSC are associated with the same GMC. 
Blue and red colors indicate Type II and Type III GMCs, respectively.
The solid line is fit to the data by the least-square method.
The Spearman's rank correlation coefficients of (a) and (b) are 0.69 and 0.70, respectively.
}\label{fig:Haysc}
\end{figure*}

\begin{figure}[htbp]
 \begin{center}
  \includegraphics[width=65mm]{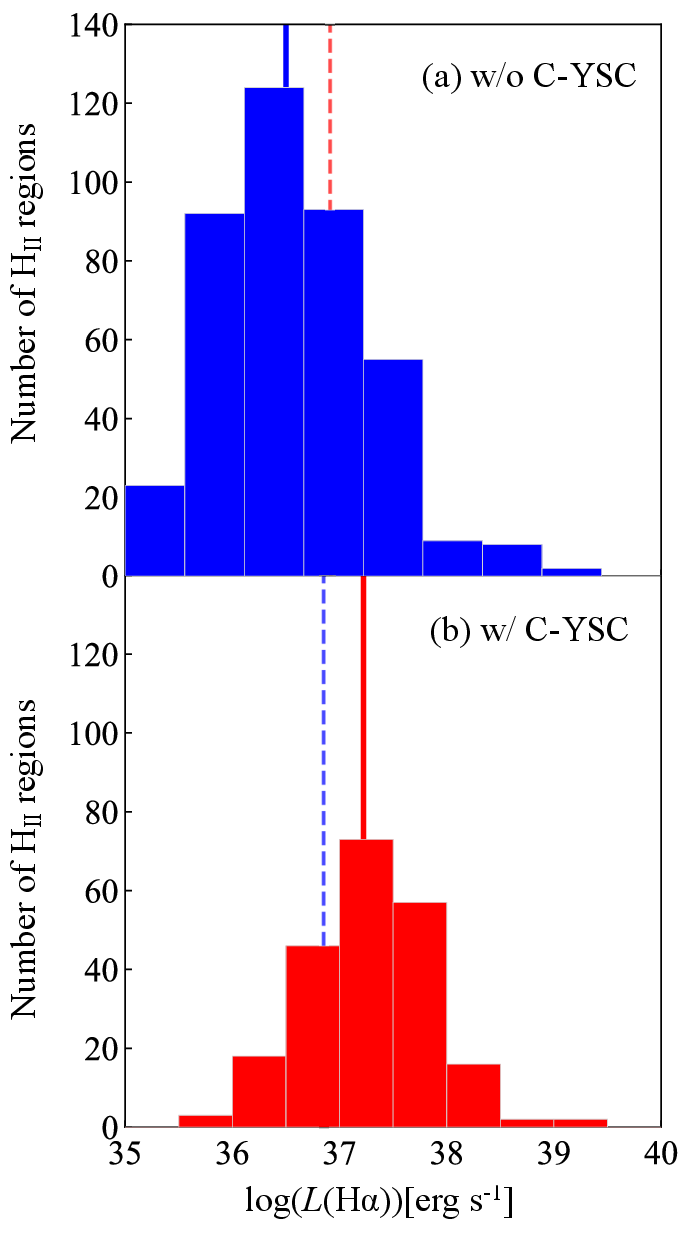}
 \end{center}
 \caption{(a) Frequency distribution of \textit{L}\,(H$\alpha$) of the H$\,${\sc ii} regions associated with GMCs without C-YSC(s). (b) Same as (a), but with C-YSC(s). Blue and red lines indicate the median values of the histograms in (a) and (b), respectively.
}\label{fig:Lha}
\end{figure}

\subsection{Physical properties of H$\,${\sc ii} regions and young stellar clusters}\label{R:yschii}
If H$\,${\sc ii} regions are formed when high-energy UV radiation from young massive clusters ionize the surrounding neutral hydrogen gas,
the H$\alpha$ luminosities of the H$\,${\sc ii} regions correlate with the physical properties of the clusters such as mass (e.g., \cite{Schombert13}, \cite{Boselli20}, \cite{Scheuermann23}).

Such a correlation between the properties of H$\,${\sc ii} regions and clusters is closely tied to the process of star formation within galaxies.
The scatter plots in Figure~\ref{fig:Haysc} show the relations between \textit{L}\,(H$\alpha$) of H$\,${\sc ii} regions and the properties of C-YSCs, specifically the total \textit{L}\,(IR) and mass. These H$\,${\sc ii} regions and C-YSCs are associated with the same GMCs.
We found that there are positive correlations between \textit{L}\,(H$\alpha$) of H$\,${\sc ii} regions and each of the total \textit{L}\,(IR) and mass of C-YSCs; the total \textit{L}\,(IR) is proportional to the \textit{L}\,(H$\alpha$) raised to the power of 1.04, and the mass is proportional to the \textit{L}\,(H$\alpha$) raised to the power of 0.91.
The Spearman's rank correlation coefficient between \textit{L}\,(H$\alpha$) and \textit{L}\,(IR) is 0.69, and that between \textit{L}\,(H$\alpha$) and mass is 0.70.
These results indicate that \textit{L}\,(H$\alpha$) of H$\,${\sc ii} regions can trace the degree of the YSC formation in GMCs, i.e., evolutionary stages of GMCs.
In the LMC, \citet{Ochsendorf16} quantitatively investigated the association of young stellar objects based on the Spitzer and Herschel infrared data, GMCs, and optical stellar clusters.
They revealed that most of Type I GMCs are not associated with high-mass star formation. \citet{Ochsendorf17} also find that the star formation rate of GMCs increases across the evolutionary sequence of GMCs from Type I to Type III proposed by \citet{Kawamura09}.
These results suggest that we can follow the general evolutionary trends of GMCs using only \textit{L}\,(H$\alpha$) although the comparison with infrared data is of course important for investigating more detailed evolutionary stages of star formation such as the earliest embedded phases with a short timescale.

\begin{table*}[h]
\tbl{Comparison of the number of GMCs at each evolutionary stage with \citet{Corbelli17}.}{%
\begin{tabular}{@{}lc@{}cc@{}}
\noalign{\vskip3pt} \hline\hline \noalign{\vskip2pt}
\multicolumn{4}{c}{This work}\\
\noalign{\vskip2pt} \hline \noalign{\vskip2pt}
Evolutionary stage & Type I & Type II & Type III \\ 
Main properties & w/o H$\,${\sc ii} regions & w/ H$\,${\sc ii} regions of \textit{L}\,(H$\alpha$) $<$ 10$^{37.5}$erg\,s$^{-1}$  & w/ H$\,${\sc ii} regions of \textit{L}\,(H$\alpha$) $>$ 10$^{37.5}$erg\,s$^{-1}$  \\ 
Number of GMCs\,\footnotemark[$a$]  & 113 (19\,$\%$)  & 361 (59\,$\%$) & 137 (22\,$\%$)\\
\noalign{\vskip2pt} \hline \noalign{\vskip2pt}
\multicolumn{4}{c}{\citet{Corbelli17}}\\
\noalign{\vskip2pt} \hline \noalign{\vskip2pt}
Evolutionary stage & Class A & Class B & Class C \\
Main properties & w/o multi-wavelength & w/ only 8\,$\mu$m, 24\,$\mu$m & w/ 8\,$\mu$m, 24\,$\mu$m, H$\alpha$, far-UV \\ 
Number of GMCs\,\footnotemark[$b$]  & 127 (27\,$\%$)  & 79 (17\,$\%$) & 268 (56\,$\%$)\\
\noalign{\vskip2pt} \hline
\end{tabular}}
\begin{tabnote}
{\hbox to 0pt{\parbox{170mm}{\footnotesize \par\noindent
\footnotemark[$a$]: Above the completeness limit, which are estimated to be 5$\times\,$10$^{4}$\,$M_{\odot}$ by referring to \citet{Engargiola03} that reported the completeness limit is about seven times larger than the lowest GMC mass (see also \cite{Muraoka23}).
\footnotemark[$b$]: Above the completeness limit of 6.3$\times\,$10$^{4}$\,$M_{\odot}$ \citep{Corbelli17}.
\par\noindent}}}
\end{tabnote}\label{table:comtype}
\end{table*}

Figure~\ref{fig:Lha} shows the frequency distributions of \textit{L}\,(H$\alpha$) for H$\,${\sc ii} regions associated with GMCs, which are separated according to their association with C-YSC(s).
There is a clear distinction in the \textit{L}\,(H$\alpha$) median between H$\,${\sc ii} regions with and without C-YSCs, with values of 7.1\,$\times$\,10$^{36}$\,erg\,s$^{-1}$ and 1.7\,$\times$\,10$^{37}$\,erg\,s$^{-1}$, respectively.
However, the difference in the frequency distribution of \textit{L}\,(H$\alpha$), between H$\,${\sc ii} regions with and without C-YSCs, is not as clear as those reported for the LMC by \citet{Yamaguchi01}.
In other words, there are low-H$\alpha$-luminosity H$\,${\sc ii} regions with C-YSCs and high-H$\alpha$-luminosity H$\,${\sc ii} regions without C-YSCs at a certain rate in M33.
The former can be potentially explained by the embedded clusters being associated with low-H$\alpha$-luminosity H$\,${\sc ii} regions, which appears as the scatter of Figure~\ref{fig:Haysc}.
The latter is partly because of the lower angular resolution of 6\farcs0 (24\,pc) of the 24\,$\mu$m data used for the identification and cataloguing of C-YSCs.
In this case, multiple neighboring clusters are likely regarded as a single C-YSC.
In our analysis, a C-YSC is determined to be associated with GMCs when the cluster's center is within the boundary of the GMCs because the \citet{Corbelli17} catalogue lacks the information on the spatial extent of C-YSCs.
Thus the decrease in the number of resolved C-YSCs decreases the chance of assigning the correspondence between GMCs, H$\,${\sc ii} regions, and C-YSCs, which eventually causes high-H$\alpha$-luminosity H$\,${\sc ii} regions without C-YSCs.

\section{Lifetime of GMCs}\label{D:lifetime}
In star formation processes, GMCs could trace a path from the quiescent phase to active star formation phase as they evolve in time.
\citet{Kawamura09} found that $\sim$70\,$\%$ of YSCs younger than 10\,Myr are associated with GMCs and considered the timescale for the GMCs in the cluster-forming phase to be 7\,Myr in the LMC.
Assuming that the star formation proceeds nearly steadily, the timescale of each evolutionary stage is considered proportional to the number of GMCs.
They estimated the timescales for each evolutionary stage from the number ratios of each type of GMC, yielding a lifetime of 26\,Myr. 
In this section, we compare the GMC classification in this study with that in \citet{Corbelli17} (Section~\ref{R:comtype}).
In addition, we examine the GMC lifetime and compare it with the previous studies in M33 (Section~\ref{R:lifetimeLMC}).

\subsection{Comparison with the classification of GMCs by Corbelli et al.(2017)}\label{R:comtype}
We consider the completeness limit of the ACA survey of $M_{\rm CO} = 5\times\,$10$^{4}$\,$M_{\odot}$ (\cite{Muraoka23}), which yields 113 (19\,$\%$) Type I GMCs, 361 (59\,$\%$) Type II GMCs, and 137 (22\,$\%$) Type III GMCs in M33.
Table~\ref{table:comtype} summarizes the comparison of the number of GMCs at each evolutionary stage in M33 between this study and \citet{Corbelli17}.
These two classifications of GMCs differ in the definition of GMC's evolutionary stages and the classification criteria.
\citet{Corbelli17} classified GMCs into three stages; inactive (Class A), embedded (Class B), and exposed (Class C) phase of high-mass star formation.
Class A GMCs show no apparent signature of high-mass star formation, Class B GMCs have only mid-infrared 8\,$\mu$m and 24\,$\mu$m emission without an optical counterpart, and Class C GMCs have multiple mid-infrared emission with H$\alpha$ and far-UV emission.
Although we differentiate between Type II and Type III covering the stages from moderate to active high-mass star-formation based on optical (H$\alpha$) emission, \citet{Corbelli17} categorized all the exposed star formations into a single stage as Class C.
As a result, Class C GMCs have the highest proportion of 56\,$\%$ among Class A, B, and C.
Meanwhile, the number of Type II and Type III GMCs depends on the classification criteria of \textit{L}\,(H$\alpha$) of H$\,${\sc ii} regions.
Both in this work and \citet{Kawamura09}, Type III GMCs account for less than 30\,$\%$ of the entire GMCs, and the most dominant phase is Type II, accounting for 50 -- 60\,$\%$.

\begin{table*}[t]
\tbl{Comparison with the GMC lifetime estimated by \citet{Corbelli17}. }{%
\begin{tabular}{@{}l@{\hspace{10mm}}c@{\hspace{28mm}}c@{\hspace{28mm}}c@{\hspace{28mm}}c@{\hspace{2.5mm}}}
\noalign{\vskip3pt} \hline\hline\noalign{\vskip2pt}
\multicolumn{5}{c}{This work}\\
\noalign{\vskip2pt}\hline
\noalign{\vskip2pt}
GMC type & Type I & Type II & Type III & Total \\ 
Number of GMCs\,\footnotemark[$a$]  & 113 (19\,$\%$)  & 361 (59\,$\%$) & 137 (22\,$\%$) & 611\\
Timescale (Myr)  & 4 & 13\,\footnotemark[$b$] & 5 & 22\\
\noalign{\vskip2pt}\hline
\noalign{\vskip2pt}
\multicolumn{5}{c}{\citet{Corbelli17}}\\
\noalign{\vskip2pt}\hline
\noalign{\vskip2pt}
GMC class & Class A & Class B & Class C & Total \\
Number of GMCs\,\footnotemark[$a$]  & 127 (27\,$\%$)  & 79 (17\,$\%$) & 268 (56\,$\%$) & 474\\
Timescale (Myr)  & 3.8 & 2.4 & 8 & 14.2\\
\noalign{\vskip2pt}\hline
\end{tabular}}
\begin{tabnote}
{\hbox to 0pt{\parbox{170mm}{\footnotesize
\par\noindent
\footnotemark[$a$]: Above the completeness limit.
\footnotemark[$b$]: The timescale of 13\,Myr for Type II GMC was originally derived in the LMC \citep{Kawamura09} and we assume that Type II GMC in M33 has the same timescale as the LMC.
\par\noindent}}}
\end{tabnote}
\label{table:GMCtimescale}
\end{table*}

We then discuss the discrepancy in the number fraction between Type I GMCs in this study (19$\%$) and Class A GMCs in \citet{Corbelli17} (27$\%$).
Through visual inspection, \citet{Corbelli17} determined the association between GMCs and multi-wavelength emissions if the emissions are highly overlapping at the center of the GMC, which is a stricter criterion than ours for the association between GMCs and H$\,${\sc ii} regions.
Based on Fig.4 in \citet{Corbelli17}, some Class A and B GMCs are lying close to H$\alpha$ emission and they are likely to be classified into Type II or Type III in this work. 
Additionally, the IRAM CO data used in \citet{Corbelli17} cover a large area of the M33 disk up to the galactocentric radius of 7.7\,kpc where it is easier to recover the low mass clouds in less crowded fields with inactive star formation.
These factors give a larger number fraction of Class A GMCs in \citet{Corbelli17} (27\,$\%$) than that of our Type I (19\,$\%$), even though we used the ACA+IRAM CO data with a higher spatial resolution, which can recover the low mass and inactive GMCs, than the IRAM CO data.
Note that we do not consider the embedded phase but it is included in our Type I phase following the definition of GMC's evolutionary stages.

\subsection{Lifetime of GMCs in M33 and its comparison with previous studies}\label{R:lifetimeLMC}
We tried estimating the GMC lifetime with reference to the previous studies of the GMC evolution in the LMC \citep{Kawamura09} without cluster information of M33.
First, the timescale for Type II GMC of 13\,Myr was originally determined in the LMC, and we assume that Type II GMC in M33 has the same timescale as the LMC.
This is because Type II GMCs are considered to be the least sensitive to the detection limit better than Type I GMCs whose typical mass ($<$\,10$^{5}$\,$M_{\odot}$) is the lowest in the three types of GMCs, and also the least sensitive to gas dissipation due to stellar feedback better than Type III GMCs which have the most active star-forming regions.
If we assume the time-steady star formation rate in M33, we can estimate the timescales of each type of GMC from the number of Type I, II, and III GMCs as shown in Table~\ref{table:comtype}.
Eventually, we obtained the timescales for each type of GMC to be 4\,Myr, 13\,Myr, and 5\,Myr, respectively, indicating the total lifetime of a GMC to be $\sim$22\,Myr.
This estimate is similar to that in the LMC derived by \citet{Kawamura09} while it is longer than that of 14.2\,Myr in M33 derived by \citet{Corbelli17}.
Note that the number of GMCs and the timescale in each type changes depending on the \textit{L}\,(H$\alpha$) threshold value for the type classification, but the GMC lifetime is almost unchanged when accounting for the possible variation of 20\,$\%$ in the \textit{L}\,(H$\alpha$) threshold (see the Appendix~\ref{app:Valtimescale}).

Table~\ref{table:GMCtimescale} summarizes the comparison with the timescale of GMCs at each evolutionary stage estimated by \citet{Corbelli17}.
The discrepancy in the GMC lifetimes can be attributed to several factors, such as the number fraction of GMCs at each evolutionary stage (see Section~\ref{R:comtype}) and the timescale estimation methods.
\citet{Corbelli17} found that YSCs associated with GMCs are younger than 8\,Myr.
They assumed that 8\,Myr is the typical age of clusters when disrupting its natal GMC, and defined 8\,Myr as the timescale of exposed star formation phase Class C.
On the other hand, in this study, we used the Type II stage as the benchmark for timescale estimation, with a timescale of 13\,Myr.
Although Class C and Type II have a similar proportion of 56$\%$ and 59$\%$, their timescales are significantly different.
This difference is directly reflected in the total GMC lifetime, resulting in longer GMC timescales in this study compared to \citet{Corbelli17}.

In a more recent work of M33, \citet{Peltonen23} found that clusters reside within their parent GMCs for a duration of 4 -- 6\,Myr, and estimated the GMC lifetime of 11 -- 15\,Myr using a 35\,pc resolution ACA CO($J$=2--1) survey data (Koch et al.\,in preparation).
They divide the GMCs into two evolutionary stages; one is where GMCs show no association with clusters, and other GMCs are at the feedback phase and associate with clusters. 
The timescale for the feedback phase of 4 -- 6\,Myr, which is comparable to our Type III GMC's timescale.
However, the observation field, the classification methods of GMCs, and the definition of GMC's evolutionary stages differ from ours, which may make a difference in the total lifetime of a GMC.
These information will be described in Koch et al.\,(in preparation), and thus we cannot make a detailed comparison in this paper.

\begin{figure*}[htbp]
 \begin{center}
  \includegraphics[width=160mm]{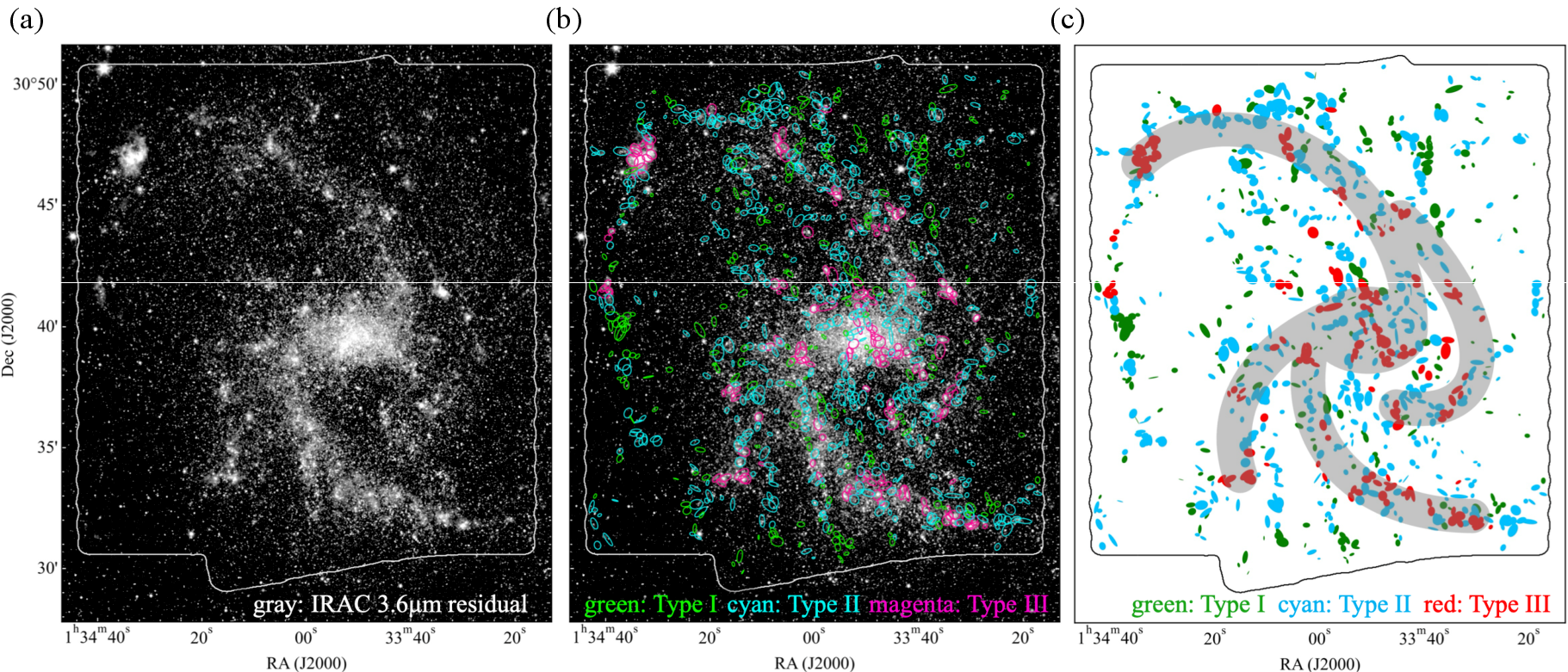}
 \end{center}
 \caption{(a) The axisymmetric model-subtracted IRAC 3.6\,$\mu$m image of M33. The spiral arm structures are clearly seen. The solid line shows the ACA observed area.
 (b) Spatial distribution of the three types of GMCs on (a). Green, cyan, and magenta ellipses represent Type I, II, and III GMCs, respectively. 
 (c) Spatial distribution of the three types of GMCs. The shaded area indicates the spiral arms and bulge traced by (a).
}\label{fig:irac3p6fitred}
\end{figure*}

\section{Discussion}\label{sec:dis}
We have demonstrated that H$\alpha$ emission can serve as a good indicator of the high-mass star formation activity within GMCs and also the GMC evolution (Section~\ref{R:ysc}).
GMCs show changes in their physical properties according to the types (Section~\ref{R:gmcprop}).
Therefore the star formation process truly transforms the GMC properties as a function of time.
Conversely, such changes in the properties between different types allow us to gain insights into the formation, evolution, and eventual dispersal or collapse of molecular clouds, and thus the validity of the type classification is confirmed.
In this section, we discuss the evolutionary process of GMCs and the effect of spiral arms on the GMC evolution (Section~\ref{D:spiralformation}), and the consistency of the GMC lifetime with the galactic dynamics (Section~\ref{D:consistency}).

\begin{figure*}[htbp]
 \begin{center}
  \includegraphics[width=160mm]{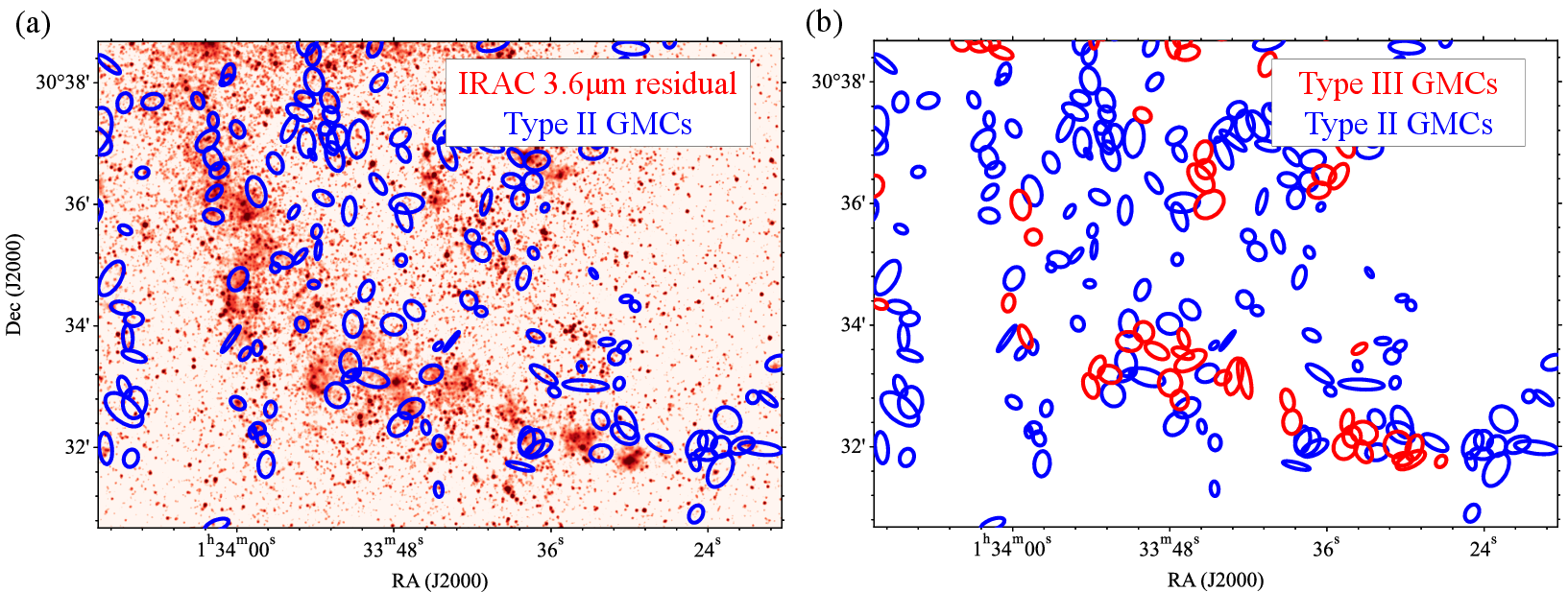}
 \end{center}
 \caption{(a) Spatial distribution of Type II GMCs in model-subtracted IRAC 3.6\,$\mu$m image around the southern spiral arm of M33.
(b) Spatial distribution of Type II and Type III GMCs represented by blue and red ellipses, respectively.
}\label{fig:typeGMCs}
\end{figure*}

\subsection{The effect of spiral arms on the GMC evolution}\label{D:spiralformation}
GMCs are considered to form and evolve by continuous mass-accretion from local surrounding H$\;${\sc i} gas as predicted by numerical studies (e.g., \cite{Hennebelle99}; Koyama \& Inutsuka \,\yearcite{Koyama00}, \yearcite{Koyama02}; \cite{Vazquez07,Goldbaum11,Inoue12,Inutsuka15}; Kobayashi et al.\,\yearcite{Kobayashi17}, \yearcite{Kobayashi18}).
Indeed, H$\;${\sc i} intensity increases from quiescent GMCs to active star-forming GMCs in the LMC \citep{Fukui09}, suggesting that H$\;${\sc i} gas accretion contributes mass growth and the evolutionary type transition of GMCs.
One of the possible mechanisms of this H$\;${\sc i} gas accretion is shock compression of H$\;${\sc i} gas by the expansion of supernova remnants.
This process expects a typical evolutionary timescale of 20 -- 60\,Myr at 10$^{6}$\,$M_{\odot}$ (Kobayashi et al.\,\yearcite{Kobayashi17}, \yearcite{Kobayashi18}).
Alternatively, the cloud-cloud collision has been recently suggested to be a dominant mechanism of high-mass star formation (see the review by \cite{Fukui21}).
In M33, the first evidence for molecular-cloud collision is found in GMC-37 \citep{Sano21}, where some molecular clouds are likely associated with supernova remnants, and the supernova feedback may cause the collision.
Galactic-scale H$\;${\sc i}-cloud collisions originating from tidal interaction with M31 also induced the GMC formation and led to the super star cluster formation in NGC~604 (e.g., \cite{Tachihara18,Muraoka20}).

As we discuss below, the spiral arm also affects the evolution of GMCs.
In the spiral arms, gas collected by the spiral potential forms GMCs and eventually high-mass stars are formed in such GMCs.
Additionally, the higher supernovae rate assists H$\,${\sc i} gas accretion onto GMCs \citep{Kobayashi18}.
M33 is one of the representative flocculent spiral galaxies, and has poorly defined arms.
In order to show the spiral structures, 
we created an axisymmetric model-subtracted image \citep{Regan94} using the Spitzer-IRAC 3.6\,$\mu$m near-infrared data, which mainly trace the old stellar population as the dominant mass component of disk galaxies (see Figure~\ref{fig:irac3p6fitred}~a, and this procedure summarized in the Appendix~\ref{app:crSpiral}).

Figure~\ref{fig:irac3p6fitred}~(b) shows the spatial distribution of the three types of GMCs on the model-subtracted IRAC 3.6\,$\mu$m image.
In Figure~\ref{fig:irac3p6fitred}~(c), we delineated the spiral arms and bulge which are roughly traced by the model-subtracted IRAC 3.6\,$\mu$m image.
We found that Type I GMCs are mainly distributed in the inter-arm, Type II GMCs are both in the arm and the inter-arm, and Type III GMCs are distributed  in the arms.
In Figure~\ref{fig:typeGMCs}, we show zoomed-in views into the southern spiral arm region, whose spiral arm structure is more prominent than the northern spiral arm (see also Figure~\ref{fig:irac3p6fitred}~a).
Type II GMCs seems distributed surrounding the arm and Type III GMCs.
These results suggest that GMCs tend to gradually evolve and move from both leading and trailing sides towards the arms, and also suggest the spiral arm plays an important role in massive star formation even in flocculent spiral galaxy M33.
We will analyze the velocity structure of CO and H$\;${\sc i} data to reveal the details of gas dynamics around spiral arms in a forthcoming paper
(Konishi et al. in preparation).

\begin{figure*}[htbp]
 \begin{center}
  \includegraphics[width=110mm]{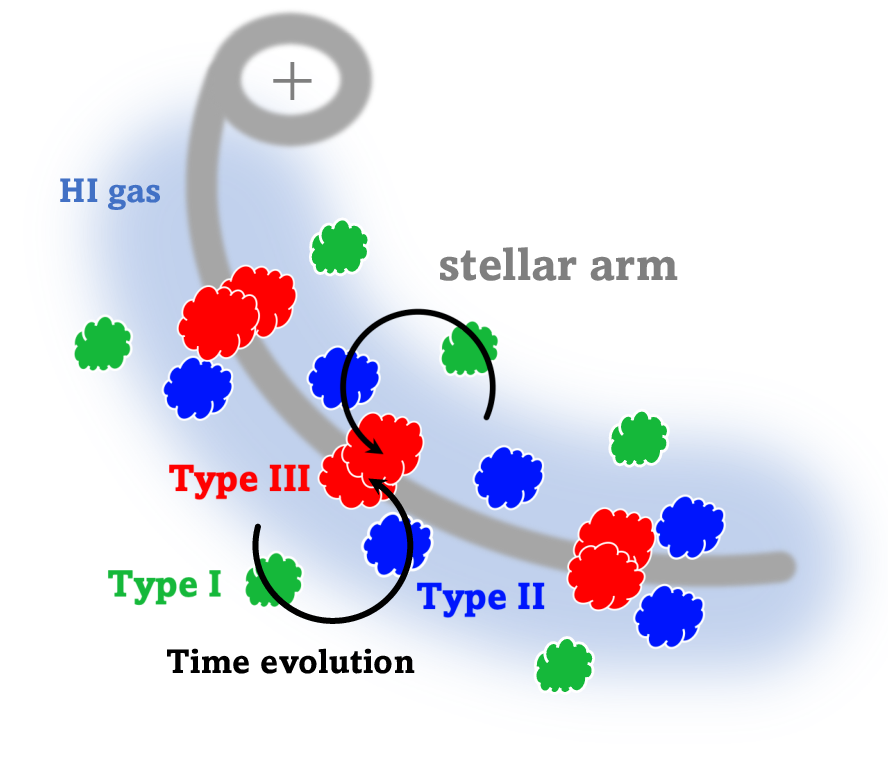}
 \end{center}
 \caption{Schematic diagram of the GMC evolution around the spiral arm of M33.
In the spiral arm, gas is collected by the spiral potential.
GMCs move from both leading and trailing sides towards the arms, and collide with each other.
Eventually, high-mass stars are formed in such colliding GMCs.
Type I GMCs distribute sparsely and mainly exist in inter-arm regions, Type II GMCs surround spiral arms, and Type III GMCs are distributed along the arms.
}\label{fig:CCCatArm}
\end{figure*}

In addition, such GMC distributions around the spiral arm suggest a likely scenario of high-mass star formation in M33; a GMC collides with other GMC(s) moving from the opposite side of the arm, and they are subsequently compressed and high-mass star formation is triggered (Figure~\ref{fig:CCCatArm}). 
This scenario may support the dynamic spiral theory (e.g., \cite{Sellwood84,Fujii11,Baba13}), which is one of the most dominant theories to explain how spiral arms form and sustain in disk galaxies.
In contrast to the traditional density wave theory (e.g., \cite{Lin64,Fujimoto68,Roberts69,Shu73}), the dynamic spiral theory has been recently developed (see the review by \cite{Dobbs14}).
The spiral arms are considered to change on the galactic rotation timescale or even more rapidly (e.g., \cite{Fujii11,Grand12,Baba13}).
The gas does not flow through a spiral arm, but rather effectively falls into the spiral potential minimum from both sides of the arm (\cite{Dobbs08,Wada11}; Baba et al.\,\yearcite{Baba16}, \yearcite{Baba17}).

Wada, Baba, and Saitoh (\yearcite{Wada11}) also argued that dynamic spiral theory might explain the complicated gas distribution such as M33 based on their hydrodynamic simulation. 
However, an observational evidence which supports the density wave theory has been reported in M33.
\citet{Tokuda20} discovered a 50\,pc scale giant molecular filament (GMF) in the GMC-16 which exists in the northern spiral arm of M33. 
The elongations of the filamentary structure are roughly perpendicular to the direction of the galactic rotation and several H$\;${\sc ii} regions are located at the downstream side relative to the filaments with offsets of $\lesssim$ 10 -- 20\,pc, suggesting that the GMFs are produced by a galactic spiral shock \citep{Fujimoto68,Roberts69,Shu73}. 
They suggested that the high-mass star formation at the southern part of GMC-16 is triggered by a collision between the GMF and a pre-existing small cloud.

In both the density wave and the dynamic spiral theories, cloud-cloud collisions are likely to occur primarily near the spiral arms where multiple molecular clouds congregate, potentially triggering the high-mass star formation.
According to recent hydrodynamic simulations of isolated galaxies, cloud-cloud collisions occur within a timescale of $\lesssim$\,5\,Myr in the dynamic spiral case (e.g., \cite{Baba17}) and 8 -- 10\,Myr in the density spiral case (e.g., \cite{Dobbs15}). 
These results indicate that the spiral arm increases the collision frequency regardless of the spiral arm dynamics \citep{Baba17}.
However, it is not clear whether collisions occur uniformly in the spiral arm of M33.
Further detailed observation towards GMCs in various environments at a sufficiently high spatial resolution to identify individual cloud-cloud collision events will lead us to well understand the mechanisms of the formation and evolution of GMCs.
Recently, \citet{Kobayashi17} calculated the time-evolution of GMC mass function, and found that cloud-cloud collision plays a smaller role in the mass growth of GMCs compared to the H$\;${\sc i} gas accretion, while the star formation triggered by cloud-cloud collisions may contribute to a few 10 percent of the total star formation \citep{Kobayashi18}.
This suggests that cloud-cloud collisions may play a role in massive star formation (e.g., transition from Type I to II and III) but not in the overall mass-growth of GMCs.
Future work with the comparison of the CO and H$\;${\sc i} data should examine the role of the local surrounding H$\;${\sc i} gas in governing the GMC evolution.

\subsection{Consistency of the GMC lifetime with galactic dynamics}\label{D:consistency}
In Section~\ref{D:spiralformation}, we suggest that GMCs move from both leading and trailing sides of the spiral arm, and then they become more active in high-mass star formation.
The closer GMCs are to the arms, the more evolved GMCs are.
This result implies that the GMC lifetime may depend on a galactic rotation time if the spiral arm is steady and follows the density wave theory.
The galactic rotation time of M33 is 100 -- 200\,Myr at galactocentric radii of 1 -- 3\,kpc estimated from the rotation curve \citep{Corbelli14}.
This timescale is considerably longer than the GMC lifetime of 22\,Myr that we derived.
In M33, observational results support both the density wave and dynamic spiral theories (see Section~\ref{D:spiralformation}).
If the spiral arm follows the dynamic spiral theory, the spiral arms are predicted to change over timescales shorter than the galactic rotation timescale (e.g., \cite{Fujii11,Grand12,Baba13}). Thus, the GMC lifetime of $\sim$\,22\,Myr is considered reasonable.

However, if we assume that the spiral arm of M33 is density-wave-like, the GMC lifetime we estimated is contradictory to that expected from the galactic rotation time.
There are some possible reasons for this inconsistency.
One reason is that we underestimate the number and timescale of Type I GMCs.
The detection limit of our CO($J$=2--1) data does not allow us to capture small molecular clouds, which yields the underestimate of Type I GMCs rather than Type II and Type III GMCs.
The actual timescale of Type I GMCs, which may be proportional to the number of them, is considered to be longer, and thus the total lifetime of GMCs also to be longer.
If this assumption holds true, it follows that more than 90\,$\%$ of Type I GMCs are not detected.
Such undetected Type I GMCs should be compact and less massive. Thus, the increasing trend in $M_{\rm CO}$ and size from Type I to Type III GMCs (see Figure~\ref{fig:GMCprop}) remain unchanged.
Nevertheless, 
we note that our results and their interpretation (see Figure~\ref{fig:CCCatArm}) are inconsistent with the idea that the spiral arm of M33 follows the density wave theory;
the observational fact that the evolved GMCs are located almost in the spiral arms while the younger GMCs are apart from the arm both in the leading side and the trailing side of the arms is consistent with the dynamic spiral theory.

Another reason is the environmental dependence of the evolutionary timescale of GMCs.
The surface density of molecular gas in the inter-arm regions is relatively low compared to the spiral arm regions, suggesting low collision frequency.
Thus GMCs in the inter-arm stay at the inactive star-forming phase for a long time.
Once GMCs reach the vicinity of spiral arms, cloud-cloud collisions are considered to be more frequent \citep{Dobbs15}, resulting in the rapid GMC evolution relative to inter-arm regions.
Hence, the closer GMCs are to the spiral arm, the shorter lifetime GMCs are considered to have.
A long GMC lifetime equivalent to the inter-arm-crossing time ($\sim$\,100\,Myr) has been proposed by the fact that a lot of GMCs are detected even in inter-arm regions (e.g., \cite{Koda09}).
\citet{Chevance20} found that molecular clouds stay in an inactive star-forming phase ($``$CO-only phase$"$) for a long time, which accounts for 75 -- 90 percent of their total lifetime.
Our Type I GMCs, which mainly exist in the inter-arm and their true lifetime may be longer, may correspond to this CO-only phase.

Note that we should consider the cloud disruption by not only stellar feedback in a later stage of high-mass star formation but also galactic dynamics.
In the grand-design spiral galaxy M51, GMCs in the downstream side of spiral arms and the inter-arm regions are dispersed by shear motion \citep{Koda09, Miyamoto14, Meidt15}, implying a short GMC lifetime in inter-arm of 20 -- 30\,Myr \citep{Meidt15}.
In the case of M33, there is a possibility of the impact of shear motion on the GMC disruption, resulting in a short GMC lifetime.

\section{Summary} \label{sec:sum}
To unravel the GMC evolution in the spiral galaxy M33, we have analyzed CO($J$=2--1) data with the ACA 7\,m array and the IRAM 30\,m at a spatial resolution of 30\,pc. 
In reference to the previous studies of GMC evolution in the LMC, we develop the classification of GMCs based on the association with H$\,${\sc ii} regions and their H$\alpha$ luminosities; Type I: associated with no H$\,${\sc ii} regions, Type II: associated with H$\,${\sc ii} regions of \textit{L}\,(H$\alpha$) $<$ 10$^{37.5}$\,erg\,s$^{-1}$, Type III: associated with H$\,${\sc ii} regions of \textit{L}\,(H$\alpha$)\,$\geqq$ 10$^{37.5}$\,erg\,s$^{-1}$, and applied this classification to M33. 
The main results are summarized as follows.

\begin{enumerate}\setlength{\parskip}{0.2cm}

  \item 
  We classified the 848\,GMCs into the three types, resulting in the number of Type I, II, and III GMCs are 224, 473, and 151, respectively. $M_{\rm CO}$, radius, $\sigma_v$, and $^{13}$CO detection rate of GMCs systematically increase from Type I to Type III GMCs.
  The $M_{\rm Vir}/M_{\rm CO}$ of Type I, II, and III GMCs are 3.1, 2.3, and 1.5, respectively. These results suggest that Type III GMCs are the densest and the closest to virial equilibrium.

  \item 
  We found remarkable differences in the randomness of the GMC distributions in the M33 disk according to their types. Type I GMCs are randomly and sparsely distributed across the disk, while the randomness of Type II GMC distribution becomes weaker than that of Type I GMCs. Type III GMCs exhibit spiral-like structures in their distribution. 

  \item 
  Type III GMCs have the strongest spatial correlation with YSCs, Type II GMCs moderate correlation, and Type I GMCs are almost uncorrelated. We confirmed that GMCs can be classified into the three types proposed by \citet{Kawamura09} based only on \textit{L}\,(H$\alpha$).
  In addition, YSCs associated with Type III GMCs are much more luminous and more massive than those with Type II GMCs. We interpret that the type classification indicates an evolutionary stage of GMCs from the quiescent phase to the most active star formation phase.

  \item Assuming that the star formation proceeds nearly steadily, the timescale of GMCs in each type is proportional to the number of them. We assumed that Type II GMC has the same timescale as the LMC and estimated the timescales of Type I, II, and III GMCs to be 4, 13, and 5\,Myr, respectively. The total GMC lifetime is $\sim$22\,Myr in M33, 
  which is consistent with that derived in the LMC while it is longer than that in the previous study of M33, 14.2\,Myr \citep{Corbelli17}.

  \item On the basis that the evolved GMCs are dominant in the spiral arms while the younger GMCs are apart from the arm both in the leading side and trailing side of the arms, we consider that a likely scenario for high-mass star formation in M33 involves cloud-cloud collisions; GMCs move from both leading and trailing sides towards the arms, and a GMC collides with other GMC(s) moving from the opposite side of the arm. These collisions lead to subsequent compression of GMCs, and finally, high-mass star formation is triggered.
  This scenario supports the dynamic spiral theory, a prominent hypothesis for elucidating the formation and persistence of spiral arms within disk galaxies.

\end{enumerate}

The type classification of GMCs based on the high-mass star formation activity elucidated the changes in the physical condition of GMCs as the high-mass star formation proceeds, which sheds light on the formation, evolution, and eventual dispersal processes of molecular clouds.
We confirm that the concept of the GMC evolution can help illuminate the galaxy's evolution including the formation of spiral arms.
The environmental dependence of the spatial distribution of GMCs at different evolutionary stages in M33 is related to galactic dynamics, especially of spiral arms.
Further investigations should focus on a quantitative correlation analysis between the spatial distribution, physical properties of molecular clouds, and galactic structures.

\begin{ack}
We would like to thank the anonymous referee for useful comments that improved the manuscript. This paper makes use of the following ALMA data: ADS/JAO ALMA\#2018.A.00058.S, ADS/JAO ALMA\#2017.1.00901.S, and ADS/JAO ALMA\#2019.1.01182.S.
ALMA is a partnership of ESO (representing its member states), NSF (USA) and NINS (Japan), together with NRC (Canada), MOST and ASIAA (Taiwan), and KASI (Republic of Korea), in cooperation with the Republic of Chile.
The Joint ALMA Observatory is operated by ESO, AUI/NRAO, and NAOJ.
This work is based on observations made with the Spitzer Space Telescope, which is operated by the Jet Propulsion Laboratory, California Institute of Technology, under a contract with NASA.
This work was supported by NAOJ ALMA Scientific Research grant Nos. 2022-22B, JSPS KAKENHI (Grant Number: 24KJ1904, 18H05440, JP21H00049, JP21K13962, 21H01136, 23KJ0322),
Tokai Pathways to Global Excellence (T-GEx), part of MEXT Strategic Professional Development Program for Young Researchers,
and the establishment of university fellowships towards the creation of science technology innovation (Grant Number: JPMJFS2138).

software: CASA \citep{CASA22}, Astropy \citep{astropy18}, APLpy (v1.1.1; \cite{Robitaille12})
\end{ack}

\appendix \label{App}
\section{M33 star cluster catalogues}\label{app:clct}
\begin{figure*}[htbp]
 \begin{center}
  \includegraphics[width=160mm]{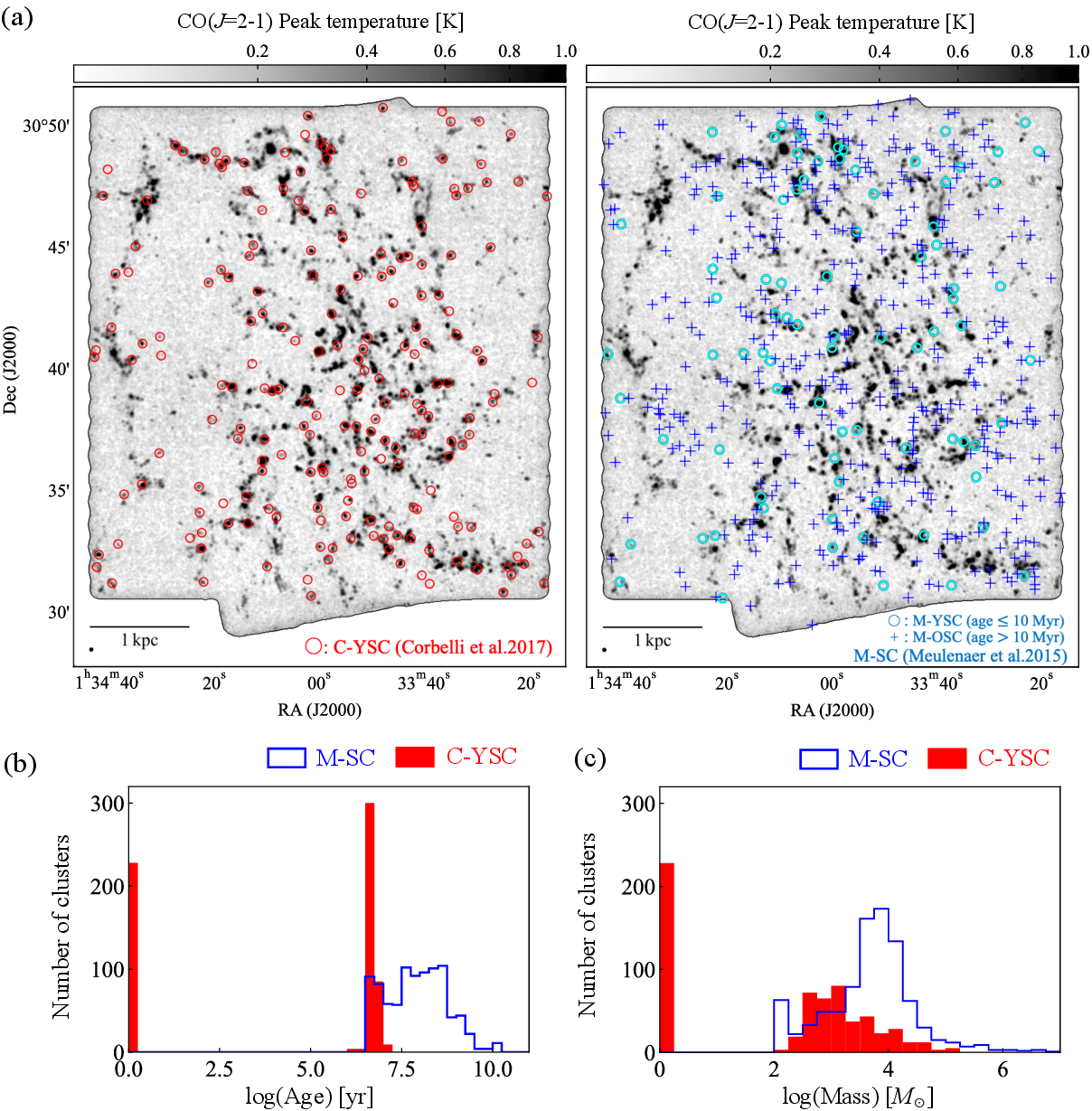}
 \end{center}
 \caption{
(a) Spatial distributions of the star clusters from two catalogues used in this work: \citet{Corbelli17} YSC (C-YSC) catalogue and \citet{Meulenaer15} star cluster (M-SC) catalogue.
Red circles indicate the positions of C-YSCs.
Cyan circles represent M-SCs which are younger than 10\,Myr (M-YSCs), and blue crosses older than 10\,Myr (M-OSCs).
The gray scale shows the $^{12}$CO peak temperature map.
The beam size of $^{12}$CO data is shown by the black ellipse in the lower left corner.
(b), (c) Histograms of age and mass of C-YSC and M-SC.
The groups whose values are zero indicate YSCs whose ages and masses have not been determined by \citet{Sharma11}.
}\label{fig:ysccatalog}
\end{figure*}

Figure~\ref{fig:ysccatalog}~(a) shows the spatial distribution of the star clusters from \citet{Meulenaer15} and \citet{Corbelli17} cluster catalogues used in this study, which are overlaid on the $^{12}$CO peak temperature map. 
These cluster catalogues cover the M33 disk beyond the ACA-observed area.

\citet{Meulenaer15} compiled a star cluster catalogue by combining three catalogues; San~Roman, Sarajedini, and Aparicio~(\yearcite{San10}) catalogue in the $\it{u', g', r', i', z'}$ photometric system of the Canada-France-Hawaii Telescope 3.6\,m, Ma\,(\yearcite{Ma12}, \yearcite{Ma13}) and \citet{Fan14} catalogues in the $\it{UBVRI}$ photometric system of the KPNO 4\,m Mayall telescope. \citet{Meulenaer15} derived physical parameters toward 910 star clusters based on the $\it{UBVRI}$ with near-infrared $\it{JHK}$ photometric systems using Two Micron All Sky Survey images. The cluster age ranges are 10$^{6.6}$ -- 10$^{10.1}$\,yr, and the mass ranges are 10$^{2.0}$ -- 10$^{6.95}$\,$M_{\odot}$ (Figure~\ref{fig:ysccatalog}~b and c).
This cluster catalogue was kindly provided by Vladas Vansevicius.

\citet{Sharma11} identified 915 mid-infrared sources using the Spitzer 24\,$\mu$m image.
They derived their physical parameters from the SED fitting using 8\,$\mu$m, 24\,$\mu$m (Spitzer), H$\alpha$ (KPNO 2.1\,m), and UV (GALEX) data and constructed a catalogue of YSC candidates. 
\citet{Corbelli17} eventually catalogued 630 YSC candidates from the list of \citet{Sharma11}.
The cluster age and mass ranges are 10$^{6.0}$ -- 10$^{7.1}$\,yr and 10$^{2.2}$ -- 10$^{5.2}$\,$M_{\odot}$, respectively (Figure~\ref{fig:ysccatalog}~b and c). 

We note that \citet{Johnson22} also reported the large star cluster catalogue consisting of 1214 objects from the Hubble Space Telescope.
The high angular resolution (0\farcs1) of this catalogue allows for the identification of individual clusters, which could provide valuable insights into the correlations between star clusters, H$\,${\sc ii} regions, and GMCs. 
However, this catalogue does not encompass the entire ACA observation field.

\section{Identified H$\,${\sc ii} regions and their properties}\label{app:idHa}
\begin{figure*}[htbp]
 \begin{center}
  \includegraphics[width=160mm]{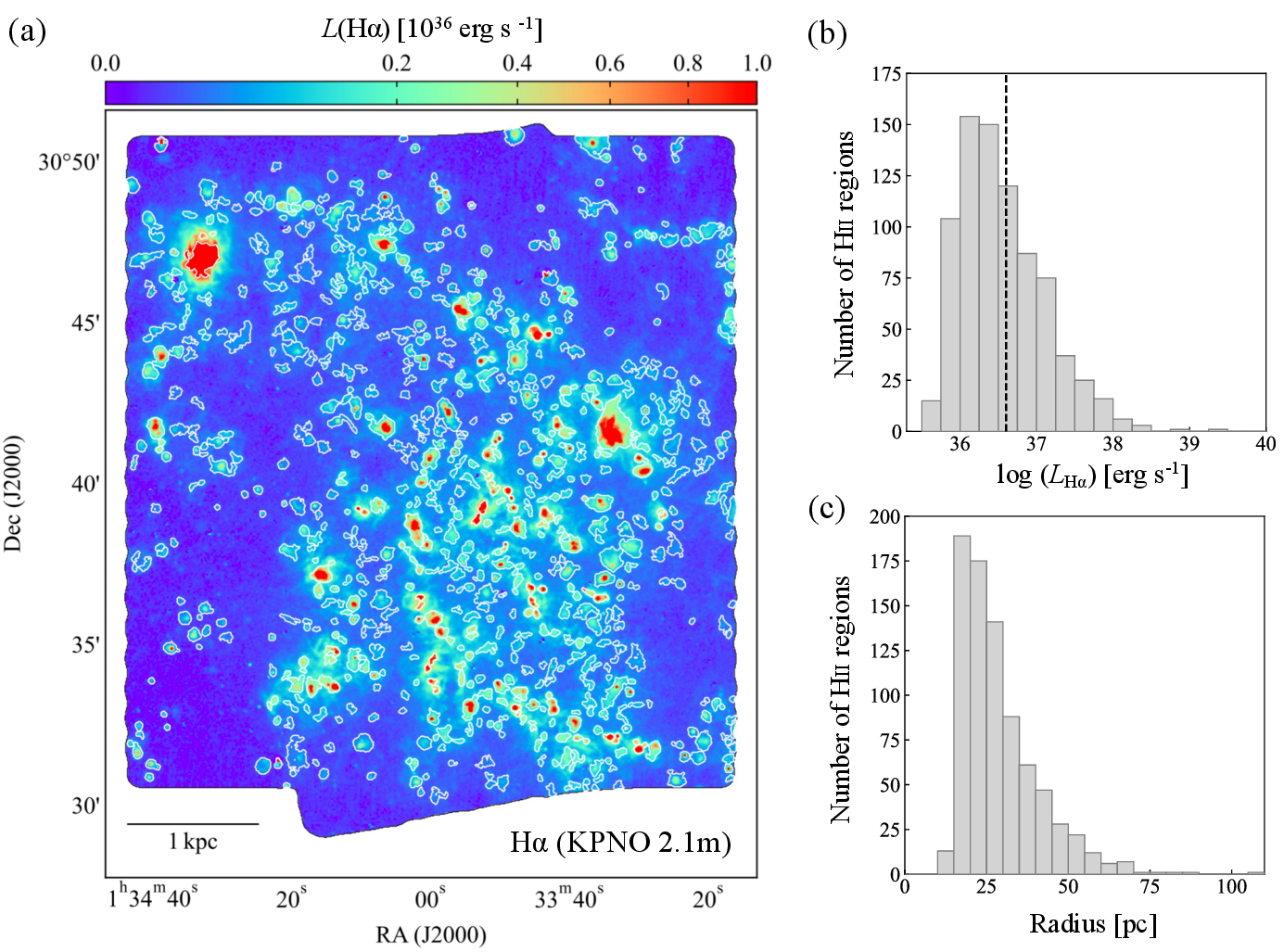}
 \end{center}
 \caption{(a) The H$\alpha$ emissions observed by KPNO 2.1\,m \citep{Hoopes00,Hoopes01}, which is cut to fit the ACA observation field. White contours denote the boundaries of the H$\alpha$ leaves identified by \texttt{astrodendro}.
 (b), (c) Frequency distributions of \textit{L}\,(H$\alpha$) and radius for identified H$\,${\sc ii} regions.
The dashed line in (b) shows the luminosity of the Orion nebula in the Milky Way, \textit{L}\,(H$\alpha$) $\sim$4\,$\times$\,10$^{36}$erg\,s$^{-1}$ \citep{Gebel68}.}
\label{fig:idHa}
\end{figure*}

Figure~\ref{fig:idHa}~(a) shows the identified H$\alpha$ structures using \texttt{astrodendro}.
We were able to roughly extract local maxima considered to be H$\,${\sc ii} regions from extended and diﬀuse H$\alpha$ emissions.
Figures~\ref{fig:idHa}~(b) and (c) show the frequency distributions of the \textit{L}\,(H$\alpha$) and the size of H$\,${\sc ii} regions, respectively.
Here we defined the effective radius 
$r_{\rm eff} = 1.91\sqrt{\sigma^{\prime}_{\rm maj}\sigma^{\prime}_{\rm min}}$, where $\sigma^{\prime}_{\rm maj}$ and $\sigma^{\prime}_{\rm min}$ are the standard deviation of the major and minor axes of the structure produced by the \texttt{astrodendro}, respectively.
The sensitivity of the H$\alpha$ data is sufficient to detect the Orion nebula with \textit{L}\,(H$\alpha$) of $\sim$\,4\,$\times$\,10$^{36}$erg\,s$^{-1}$ \citep{Gebel68}.
In Figure~\ref{fig:idHa}~(c), there is a blank space below the radius of about 15\,pc due to the thresholds for the identification of H$\,${\sc ii} regions.

We estimate uncertainties of the \textit{L}\,(H$\alpha$) and size of H$\,${\sc ii} regions using a bootstrap method following \citet{Rosolowsky06}.
This method is the process of randomly resampling data points \textit{N} times from each original H$\,${\sc ii} region that has \textit{N} data points and generating a trial structure.
We obtained the uncertainties by generating 10000 trial structures for each original H$\,${\sc ii} region and calculating the standard deviation of their properties.
The uncertainty of \textit{L}\,(H$\alpha$) is 20\,$\%$ at most, and about 80\,$\%$ of H$\,${\sc ii} regions have the uncertainty of less than 5\,$\%$.
The uncertainties of size is around 5\,$\%$.

\section{Effects of the changes of \textit{L}\,(H$\alpha$) threshold on the GMC lifetime}\label{app:Valtimescale}
The number of GMCs and the timescale in each type changes depending on the \textit{L}\,(H$\alpha$) threshold for the type classification.
We estimate the effects of the variation in \textit{L}\,(H$\alpha$) threshold on the GMC lifetime by assuming the \textit{L}\,(H$\alpha$) variation by 20\,$\%$ derived in the Appendix~\ref{app:idHa}.
If we raise the \textit{L}\,(H$\alpha$) threshold by 20\,$\%$, the number of Type III GMCs decreases and that of Type II GMCs increases.
Thus the timescale in Type III GMCs becomes shorter, yielding the total lifetime of a GMC of 21\,Myr using the method based on the LMC study (\cite{Kawamura09}, and see Section~\ref{R:lifetimeLMC}).
Conversely, if we lower the \textit{L}\,(H$\alpha$) threshold by 20\,$\%$, Type III GMCs account for a higher proportion and have a longer timescale.
This yields the GMC lifetime of 24\,Myr. 
Thus the GMC lifetime is almost unchanged from that of 22\,Myr derived in Section~\ref{R:lifetimeLMC} even though we consider the possible variation in the \textit{L}\,(H$\alpha$) threshold.

\begin{figure*}[htbp]
 \begin{center}
  \includegraphics[width=160mm]{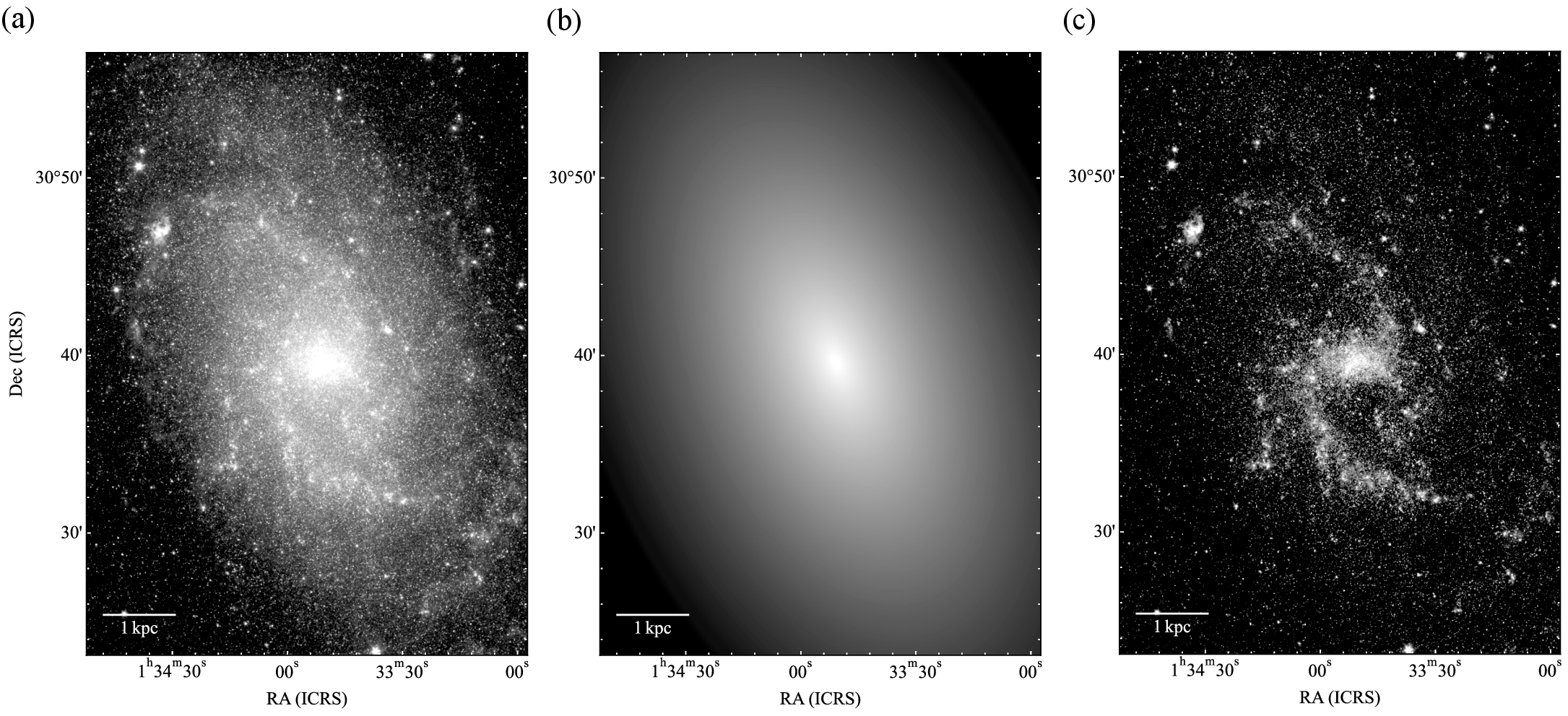}
 \end{center}
 \caption{(a) Original Spitzer-IRAC 3.6\,$\mu$m data \citep{Dale09} of M33. (b) Best fitting axisymmetric model disk of M33 from the IRAC 3.6\,$\mu$m data created by equation~(\ref{eq:sersic}) with fitting range of $I(r)$\,$<$\,1 MJy sr$^{-1}$.
(c) Model-subtracted IRAC 3.6\,$\mu$m image of M33. The spiral arm structures are seen more clearly than (a). 
}\label{fig:irac3p6fit}
\end{figure*}

\begin{figure}[htbp]
 \begin{center}
  \includegraphics[width=80mm]{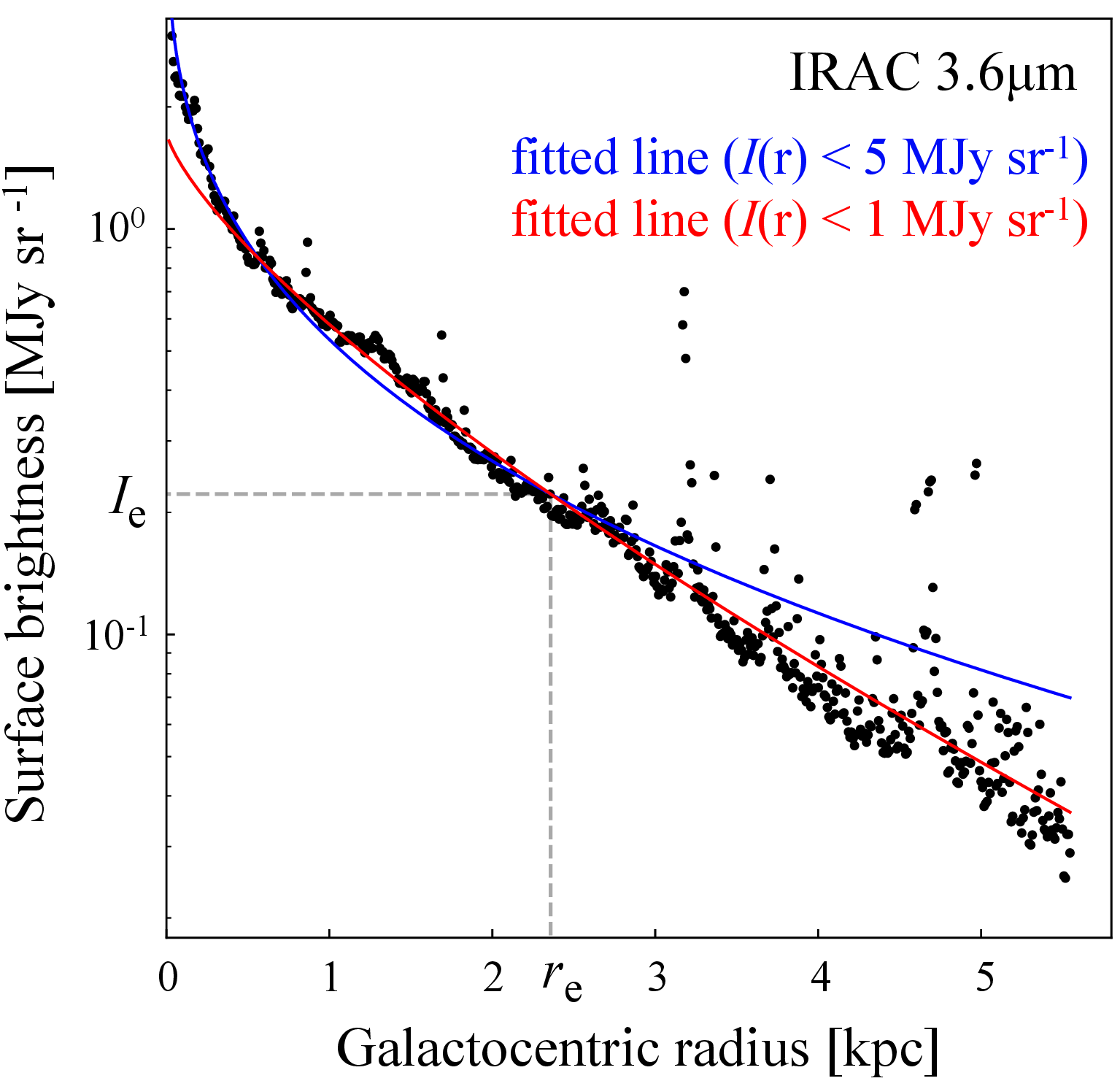}
 \end{center}
 \caption{Radial profile of 3.6\,$\mu$m surface brightness in M33. 
Red and blue lines indicate the best-fit line using the equation~(\ref{eq:sersic}). 
}\label{fig:fitplot}
\end{figure}

\section{Spiral arm structure traced by near-infrared bands}\label{app:crSpiral}
In this section, we present the procedure that create an axisymmetric model-subtracted image of M33. Firstly, we create an axisymmetric model disk of M33 \citep{Regan94} from the IRAC 3.6\,$\mu$m data (Figure~\ref{fig:irac3p6fit}~a).
\citet{Vaucouleurs48} found that surface brightness \textit{I} especially for elliptical galaxies decreases in proportion to galactocentric radius \textit{r}$^{1/4}$ ($``$de Vaucouleurs model$"$).
Besides, S\'{e}rsic\,(\yearcite{Sersic63}, \yearcite{Sersic68})
suggested the approximate equation which expresses well the surface brightness function of various galaxies, known as the S\'{e}rsic model as follows:
\begin{eqnarray}\label{eq:sersic}
I(r) = I_{e}\,\times\,\exp[-k\{(\frac{r}{r_{e}})^{\frac{1}{n}}\}-1]   
\end{eqnarray}
where \textit{r}$_{e}$ is the effective half-light radius and \textit{I}$_{e}$ is the surface brightness at \textit{r}$_{e}$.
The S\'{e}rsic index \textit{n} reflects the degree of central concentration of brightness. When \textit{n} = 4, equation~(\ref{eq:sersic}) corresponds to the de Vaucouleurs model. 
The constant $k$ is treated as a free parameter.
We calculate the average surface brightness $I(r)$ in annular elliptical rings with a width of 3\farcs6 assuming an inclination angle of 55$^{\circ}$ and a position angle of 21$^{\circ}$ \citep{Koch18}.
Figure~\ref{fig:irac3p6fit}~(b) shows the obtained model disk fitted by equation~(\ref{eq:sersic}), where the S\'{e}rsic index is \textit{n} = 1.4.
We limited the fitting range to $I(r)$\,$<$\,1 MJy\,sr$^{-1}$ because the $I(r)$ excess in the galactic center degrades the fitting accuracy (Figure~\ref{fig:fitplot}).
Actually, the fitted surface brightness seems to be overestimated compared to the raw value in the range of \textit{r} $>$ $\sim$3\,kpc when the fitting range is set to $I(r)$\,$<$\,5\,MJy\,sr$^{-1}$.

Secondly, we multiplied the model disk value by 0.7 so that the value in the weak 3.6\,$\mu$m emission regions such as inter-arm becomes around zero after the subtraction.
We subtracted the modified model disk from the original disk.
Eventually, we obtained the residual image (Figure~\ref{fig:irac3p6fit}~c, see also the similar procedure by \cite{Regan94}).
The two spiral structures in the north and south are clearly seen.
We consider that this residual map traces the spiral arm structure.


\begin{thebibliography}{}
\bibitem[Astropy Collaboration et al.(2018)]{astropy18} Astropy Collaboration., et al. 2018, \aj, 156, 123

\bibitem[Baba et al.(2013)]{Baba13} Baba, J., Saitoh, T. R., \& Wada, K. 2013, \apj, 763, 46

\bibitem[Baba et al.(2016)]{Baba16} Baba, J., Morokuma-Matsui, K., Miyamoto, Y., Egusa, F., \& Kuno, N. 2016, \mnras, 460, 2472

\bibitem[Baba et al.(2017)]{Baba17} Baba, J., Morokuma-Matsui, K., \& Saitoh, T.R. 2017, \mnras, 464, 246

\bibitem[Bica et al.(1996)]{Bica96}Bica, E., Claria, J. J., Dottori, H., Santos, J. F. C., Jr., \& Piatti, A. E. 1996, \apjs, 102, 57

\bibitem[Bolatto et al.(1999)]{Bolatto99}Bolatto, A. D., Jackson, J. M., \& Ingalls, J. G. 1999, \apj, 513, 275

\bibitem[Bolatto et al.(2013)]{Bolatto13}Bolatto, A. D., Wolfire, M., \& Leroy, A. K. 2013, ARA\&A, 51, 207

\bibitem[Boselli et al.(2020)]{Boselli20}Boselli, A., et al. 2020, \aap, 634, L1

\bibitem[CASA Team et al.(2022)]{CASA22} CASA Team et al. 2022, \pasp, 134, 114501

\bibitem[Chevance et al.(2020)]{Chevance20} Chevance, M., et al. 2020, \mnras, 493, 2872

\bibitem[Chen et al.(2015)]{Chen15} Chen, H., Gao, Y., Braine, J., \& Gu, Q. 2015, \apj, 810, 140

\bibitem[Chen et al.(2017)]{Chen17} Chen, H., Braine, J., Gao, Y., Koda, J., \& Gu, Q. 2017, \apj, 836, 101

\bibitem[Colombo et al.(2014)]{Colombo14} Colombo, D., et al. 2014, \apj, 784, 3

\bibitem[Corbelli et al.(2014)]{Corbelli14} Corbelli, E., Thilker, D., Zibetti, S., Giovanardi, C., \& Salucci, P. 2014, \aap, 572, A23

\bibitem[Corbelli et al.(2017)]{Corbelli17} Corbelli, E., et al. 2017, \aap, 601, A146

\bibitem[Cosens et al.(2022)]{Cosens22} Cosens, M, et al. 2022, \apj, 929, 74

\bibitem[Dale et al.(2009)]{Dale09} Dale D. A., et al. 2009, \apj, 703, 517

\bibitem[Della Bruna et al.(2020)]{Della20} Della Bruna, L., et al. 2020, \aap, 635, A134

\bibitem[Demachi et al.(2023)]{Demachi23} Demachi, F., et al. 2023, arXiv:2305.19192

\bibitem[de Meulenaer et al.(2015)]{Meulenaer15} de Meulenaer, P., Narbutis, D., Mineikis, T., \& Vansevi\v{c}ius, V. 2015, \aap, 581, A111

\bibitem[de Vaucouleurs (1948)]{Vaucouleurs48} de Vaucouleurs, G. 1948, Annales d’Astrophysique, 11, 247

\bibitem[Dobashi et al.(1994)]{Dobashi94} Dobashi, K., Bernard, J.-Ph., Yonekura, Y., \& Fukui, Y. 1994, \apjs, 95, 419

\bibitem[Dobashi et al.(1996)]{Dobashi96} Dobashi, K., Bernard, J.-Ph., \& Fukui, Y. 1996, \apj, 466, 282

\bibitem[Dobbs \& Bonnell(2008)]{Dobbs08} Dobbs, C. L., \& Bonnell I. A. 2008, \mnras, 385, 1893

\bibitem[Dobbs \& Baba(2014)]{Dobbs14} Dobbs, C., \& Baba, J. 2014, PASA, 31, e035

\bibitem[Dobbs et al.(2015)]{Dobbs15} Dobbs, C. L., Pringle, J. E., \& Duarte-Cabral, A. 2015, \mnras, 446, 3608

\bibitem[Druard et al.(2014)]{Druard14} Druard, C., et al. 2014, \aap, 567, A118

\bibitem[Engargiola et al.(2003)]{Engargiola03} Engargiola, G., Plambeck, R. L., Rosolowsky, E., \& Blitz, L. 2003, \apjs, 149, 343.

\bibitem[Fan \& de Grijs(2014)]{Fan14} Fan, Z., \& de Grijs, R. 2014, \apjs, 211, 22

\bibitem[Freedman et al.(1991)]{Freedman91}Freedman, W. L., Wilson, C. D., \& Madore, B. F. 1991, \apj, 372, 455

\bibitem[Fujii et al.(2011)]{Fujii11} Fujii, M. S., Baba, J., Saitoh, T. R., Makino, J., Kokubo, E., \& Wada, K. 2011, \apj, 730, 109

\bibitem[Fujimoto(1968)]{Fujimoto68} Fujimoto, M., 1968, in IAU Symp. 29, Non-stable Phenomena in Galaxies (Cambridge: Cambridge Univ. Press), 453

\bibitem[Fukui et al.(1999)]{Fukui99} Fukui, Y., et al. 1999, \pasj, 51, 745

\bibitem[Fukui et al.(2008)]{Fukui08} Fukui, Y., et al. 2008, \apjs, 178, 56

\bibitem[Fukui et al.(2009)]{Fukui09} Fukui, Y., et al.\ 2009, \apj, 705, 144

\bibitem[Fukui et al.(2021)]{Fukui21} Fukui, Y., Habe, A., Inoue, T., Enokiya, R., \& Tachihara, K. 2021, \pasj, 73, S1

\bibitem[Gao \& Solomon(2004a)]{Gao04a} Gao, Y., \& Solomon, P. M. 2004a, \apjs, 152, 63

\bibitem[Gao \& Solomon(2004b)]{Gao04b} Gao, Y., \& Solomon, P. M. 2004b, \apj, 606, 271

\bibitem[Gebel(1968)]{Gebel68} Gebel, W. L. 1968, \apj, 153, 743

\bibitem[Goldbaum et al.(2011)]{Goldbaum11}Goldbaum, N. J., Krumholz, M. R., Matzner, C. D., \& McKee, C. F. 2011, \apj, 738, 101

\bibitem[Grand et al.(2012)]{Grand12} Grand, R. J. J., Kawata, D., \& Cropper, M. 2012, \mnras, 426, 167

\bibitem[Gratier et al.(2012)]{Gratier12} Gratier, P., et al. 2012, \aap, 542, A108

\bibitem[Gratier et al.(2017)]{Gratier17} Gratier, P., et al. 2017, \aap, 600, A27

\bibitem[Haffner et al.(2009)]{Haffner09} Haffner L. M., et al. 2009, Reviews of Modern Physics, 81, 969

\bibitem[Hennebelle \& P{\'e}rault(1999)]{Hennebelle99} Hennebelle, P., \& P{\'e}rault, M. 1999, \aap, 351, 309

\bibitem[Hirota et al.(2011)]{Hirota11} Hirota, A., Kuno, N., Sato, N., Nakanishi, H., Tosaki, T., \& Sorai, K. 2011, \apj, 737, 40

\bibitem[Hoopes \& Walterbos(2000)]{Hoopes00} Hoopes, C. G., \& Walterbos, R. A. M. 2000, \apj, 541, 597

\bibitem[Hoopes et al.(2001)]{Hoopes01} Hoopes, C. G., Walterbos, R. A. M., \& Bothun, G. D. 2001, \apj, 559, 878

\bibitem[Hunter et al.(1996)]{Hunter96} Hunter, D. A., Baum, W. A., O’Neil, E. J., \& Lynds, R. 1996 \apj, 456, 174

\bibitem[Inoue \& Inutsuka(2012)]{Inoue12}Inoue, T., \& Inutsuka, S.-I. 2012, \apj, 759, 35

\bibitem[Inutsuka et al.(2015)]{Inutsuka15}Inutsuka, S.-I., Inoue, T., Iwasaki, K., \& Hosokawa, T. 2015, \aap, 580, A49

\bibitem[Johnson et al.(2022)]{Johnson22} Johnson L. C., et al., 2022, \apj, 938, 81

\bibitem[Kawamura et al.(2009)]{Kawamura09} Kawamura, A., et al.\ 2009, \apjs, 184, 1

\bibitem[Kennicutt \& Hodge(1986)]{Kennicutt86} Kennicutt, R. C., \& Hodge, P. W. 1986, \apj, 306, 130

\bibitem[Kobayashi et al.(2017)]{Kobayashi17}Kobayashi, M. I. N., Inutsuka, S.-I., Kobayashi, H., \& Hasegawa, K. 2017, \apj, 836, 175

\bibitem[Kobayashi et al.(2018)]{Kobayashi18}Kobayashi, M. I. N., Kobayashi, H., Inutsuka, S.-I., \& Fukui Y. 2018, \pasj, 70, S59

\bibitem[Koch et al.(2018)]{Koch18} Koch, E. W., et al.\ 2018, \mnras, 479, 2505

\bibitem[Koda et al.(2009)]{Koda09} Koda J., et al., 2009, \apj, 700, L132

\bibitem[Kondo et al.(2021)]{Kondo21} Kondo, H., et al. 2021, \apj, 912, 66

\bibitem[Koyama \& Inutsuka(2000)]{Koyama00} Koyama, H., \& Inutsuka, S.-I. 2000, \apj, 532, 980

\bibitem[Koyama \& Inutsuka(2002)]{Koyama02} Koyama, H., \& Inutsuka, S.-I. 2002, \apj, 564, L97

\bibitem[Kurt \& Dufour(1998)]{Kurt98} Kurt, C. M., \&  Dufour, R. J. 1998, RMxAC, 7, 202

\bibitem[Lada et al.(2010)]{Lada10} Lada, C. J., Lombardi, M., \& Alves, J. F. 2010, \apj, 724, 687

\bibitem[Leroy et al.(2021)]{Leroy21} Leroy, A. K., et al. 2021, \apjs, 257, 43

\bibitem[Lin \& Shu(1964)]{Lin64} Lin C. C., \& Shu F. H. 1964, \apj, 140, 646

\bibitem[Ma(2012)]{Ma12} Ma, J. 2012, \aj, 144, 41

\bibitem[Ma(2013)]{Ma13} Ma, J. 2013, \aj, 145, 88

\bibitem[Maddalena \& Thaddeus(1985)]{Maddalena85} Maddalena, R. J., \& Thaddeus, P. 1985, \apj, 294, 231

\bibitem[Maloney \& Black(1988)]{Maloney88} Maloney, P., \& Black, J. H. 1988, \apj, 325, 389

\bibitem[Malumuth et al.(1996)]{Malumuth96} Malumuth, E. M., Waller, W. H., \& Parker, J. W. 1996, \aj, 111, 1128

\bibitem[McLeod et al.(2021)]{McLeod21} McLeod, A. F., et al. 2021, \mnras, 508, 5425

\bibitem[Meidt et al.(2015)]{Meidt15} Meidt, S. E., et al. 2015, \apj, 806, 72

\bibitem[Miura et al.(2012)]{Miura12} Miura, R. E., et al. 2012, \apj, 761, 37 

\bibitem[Miyamoto et al.(2014)]{Miyamoto14} Miyamoto, Y., Nakai, N., \& Kuno, N. 2014, \pasj, 66, 36

\bibitem[Muraoka et al.(2009a)]{Muraoka09a} Muraoka, K., et al. 2009a, \pasj, 61, 163

\bibitem[Muraoka et al.(2009b)]{Muraoka09b} Muraoka, K., et al. 2009b, \apj, 706, 1213

\bibitem[Muraoka et al.(2020)]{Muraoka20} Muraoka, K., et al. 2020, \apj, 903, 94

\bibitem[Muraoka et al.(2023)]{Muraoka23} Muraoka, K., et al. 2023, \apj, 953, 164

\bibitem[Nishimura et al.(2015)]{Nishimura15} Nishimura, A., et al. 2015, \apjs, 216, 18

\bibitem[Ochsendorf et al.(2016)]{Ochsendorf16} Ochsendorf, B. B., Meixner, M., Chastenet, J., Tielens, A. G. G. M., \& Roman-Duval, J. 2016, \apj, 832, 43

\bibitem[Ochsendorf et al.(2017)]{Ochsendorf17} Ochsendorf, B. B., Meixner, M., Roman-Duval, J., Rahman, M., \& Evans, N. J., II 2017, \apj, 841, 109

\bibitem[Onodera et al.(2010)]{Onodera10} Onodera, S., et al. 2010, \apjl, 722, L127

\bibitem[Pan et al.(2022)]{Pan22} Pan, H. A., et al. 2022, \apj, 927, 9

\bibitem[Peltonen et al.(2023)]{Peltonen23} Peltonen, J., et al. 2023, \mnras, 522, 6137

\bibitem[Querejeta et al.(2019)]{Querejeta19} Querejeta, M., et al. 2019, \aap, 625, A19

\bibitem[Regan \& Vogel(1994)]{Regan94} Regan M. W., \& Vogel S. N. 1994, \apj, 434, 536

\bibitem[Roberts(1969)]{Roberts69} Roberts, W. W. 1969, \apj, 158, 123

\bibitem[Robitaille \& Bressert(2012)]{Robitaille12}Robitaille, T., \& Bressert, E. 2012, APLpy: Astronomical Plotting Library in Python, ascl:1208.017

\bibitem[Rosolowsky \& Leroy(2006)]{Rosolowsky06} Rosolowsky, E., \& Leroy, A. 2006, \pasp, 118, 590

\bibitem[Rosolowsky et al.(2008)]{Rosolowsky08} Rosolowsky, E. W., Pineda, J. E., Kauffmann, J., \& Goodman, A. A. 2008, \apj, 679, 1338

\bibitem[Rosolowsky \& Simon(2008)]{Simon08}Rosolowsky, E., \& Simon, J. D. 2008, \apj, 675, 1213

\bibitem[Rosolowsky et al.(2021)]{Rosolowsky21} Rosolowsky, E. W., et al. 2021, \mnras, 502, 1218

\bibitem[Sano et al.(2021)]{Sano21} Sano, H., et al. 2021, \pasj, 73, S62

\bibitem[San Roman et al.(2010)]{San10} San Roman, I., Sarajedini, A., \& Aparicio, A. 2010, \apj, 720, 1674

\bibitem[Scheuermann et al.(2023)]{Scheuermann23} Scheuermann F., et al., 2023, \mnras, 522, 2369

\bibitem[Schinnerer et al.(2013)]{Schinnerer13} Schinnerer, E., et al. 2013, \apj, 779, 42

\bibitem[Schinnerer et al.(2019)]{Schinnerer19} Schinnerer, E., et al. 2019, \apj, 887, 49

\bibitem[Schombert et al.(2013)]{Schombert13} Schombert, J., McGaugh, S., \&  Maciel, T. \aj, 146, 41

\bibitem[Schruba et al.(2010)]{Schruba10} Schruba, A., Leroy, A. K., Walter, F., Sandstrom, K., \& Rosolowsky, E. 2010, \apj, 722, 1699

\bibitem[Sellwood \& Carlber(1984)]{Sellwood84} Sellwood, J. A., \& Carlberg, R. G. 1984, \apj, 282, 61

\bibitem[S\'{e}rsic(1963)]{Sersic63} S\'{e}rsic, J. L. 1963, Bolet\'{i}n de la Asociaci\'{o}n Argentina de Astronom\'{i}a, 6, 41

\bibitem[S\'{e}rsic(1968)]{Sersic68} S\'{e}rsic, J. L. 1968, Atlas de Galaxias Australes (Cordoba: Observatorio Astronomico)

\bibitem[Sharma et al.(2011)]{Sharma11} Sharma, S., Corbelli, E., Giovanardi, C., Hunt, L. K., \& Palla, F. 2011, \aap, 534, A96

\bibitem[Shimajiri et al.(2017)]{Shimajiri17} Shimajiri, Y., et al. 2017, \aap, 604, A74

\bibitem[Shu et al.(1973)]{Shu73} Shu, F. H., Milione, V., \& Roberts, W. W., Jr. 1973, \apj, 183, 819

\bibitem[Solomon et al.(1987)]{solomon1987} Solomon, P. M., Rivolo, A. R., Barrett, J., \& Yahil, A. 1987, \apj, 319, 730

\bibitem[Skrutskie et al.(2006)]{Skrutskie06} Skrutskie, M. F., et al. 2006, \aj, 131, 1163

\bibitem[Tachihara et al.(2018)]{Tachihara18} Tachihara, K., Gratier, P., Sano, H., Tsuge, K.,  Miura, R. E., Muraoka, K., \& Fukui, Y. 2018, \pasj, 70, S52

\bibitem[Tokuda et al.(2020)]{Tokuda20} Tokuda, K., et al. 2020, \apj, 896, 36

\bibitem[Tokuda et al.(2021)]{Tokuda21} Tokuda, K., et al. 2021, \apj, 922, 171

\bibitem[Tokuda et al.(2023)]{Tokuda23} Tokuda, K., et al. 2023, \apj, 955, 52

\bibitem[Torii et al.(2019)]{Torii19} Torii, K., et al. 2019, \pasj, 71, S2

\bibitem[V{\'a}zquez-Semadeni et al.(2007)]{Vazquez07} V{\'a}zquez-Semadeni, E., G{\'o}mez, G.~C., Jappsen, A.~K., Ballesteros-Paredes, J., Gonz{\'a}lez, R. F., \& Klessen, R. S. 2007, \apj, 657, 870

\bibitem[Wada et al.(2011)]{Wada11} Wada, K., Baba, J., \& Saitoh, T. R. 2011, \apj, 735, 1

\bibitem[Wall et al.(2016)]{Wall16}Wall, W. F., et al. 2016, \mnras, 459, 1440

\bibitem[Weilbacher et al.(2018)]{Weilbacher18} Weilbacher, P. M., et al. 2018, \aap, 611, A95

\bibitem[Wilson \& Matthews(1995)]{Wilson95} Wilson, C. D., \& Matthews, B. C. 1995, \apj, 455, 125

\bibitem[Wong et al.(2019)]{Wong19} Wong, T., et al. 2019, \apj, 885, 50

\bibitem[Yajima et al.(2021)]{yajima2021} Yajima, Y., et al. 2021, \pasj, 73, 257

\bibitem[Yamaguchi et al.(2001)]{Yamaguchi01} Yamaguchi, R., et al. 2001, \pasj, 53, 985

\bibitem[Zaragoza-Cardiel et al.(2015)]{Zaragoza15} Zaragoza-Cardiel, J., et al.\ 2015, \mnras, 451, 1307

\bibitem[Zakardjian et al.(2023)]{Zakardjian23} Zakardjian, A., et al. 2023, \aap, 678, A171

\end{thebibliography}
\end{document}